\documentclass[journal,twoside]{IEEEtran}
\usepackage {multirow}
\usepackage {setspace}

\usepackage{cite}

\ifCLASSINFOpdf
  \usepackage[pdftex]{graphicx}
\else
   \usepackage[dvips]{graphicx}
   \DeclareGraphicsExtensions{.eps}
\fi

\usepackage{bm}

\usepackage{amsmath}
\usepackage{amsfonts}
\usepackage{amssymb}
\usepackage{amsthm}
\newtheorem{observation}{Observation}
\theoremstyle{remark}
\newtheorem*{remark}{Remark}
\usepackage{stfloats}
\usepackage{bm}
\usepackage{caption}
\usepackage{algorithmic}
\usepackage{algorithm}

\usepackage{amsmath}

\usepackage{array}

\ifCLASSOPTIONcompsoc
  \usepackage[caption=false,font=normalsize,labelfont=sf,textfont=sf]{subfig}
\else
  \usepackage[caption=false,font=footnotesize]{subfig}
\fi
\usepackage{bbm}
\usepackage {color}
\begin{document}
%
\title{Adaptive Processor Frequency Adjustment for Mobile Edge Computing with Intermittent Energy Supply}
%
%
%

 \author{Tiansheng~Huang,
         Weiwei~Lin,
         ~Xiaobin~Hong,
         ~Xiumin~Wang,
         ~Qingbo~Wu,
         ~Rui~Li,
         ~Ching-Hsien~Hsu, 
         and ~Albert~Y. Zomaya~, \IEEEmembership{Fellow,~IEEE}
%

 \IEEEcompsocitemizethanks{
\IEEEcompsocthanksitem  The authors would like to thank the anonymous reviewers for their constructive comments. Special thanks are due to Dr. Minxian Xu of Shenzhen Institutes of Advanced Technology, Chinese Academy of Sciences, for his constructive advice on our experiment setup. This work is supported by Key-Area Research and Development Program of Guangdong Province(2021B0101420002), National Natural Science Foundation of China (62072187, 61872084),  Guangzhou Science and Technology Program key projects (202007040002), Guangdong Major Project of Basic and Applied Basic Research(2019B030302002), and Guangzhou Development Zone Science and Technology (2020GH10). \textit{(Corresponding authors: Weiwei Lin; Xiumin Wang.)	}
\IEEEcompsocthanksitem T. Huang, W. Lin, and  X. Wang are with the School of Computer Science and Engineering, South China University of Technology, China. Email: tianshenghuangscut@gmail.com, linww@scut.edu.cn, xmwang@scut.edu.cn.
 \IEEEcompsocthanksitem X. Hong is with the School of Mechanical and Automotive Engineering, South China University of Technology, China. Email: mexbhong@scut.edu.cn.
\IEEEcompsocthanksitem CH. Hsu is with the Department of Computer Science and Information Engineering, Asia University, Taichung, Taiwan and with the Department of Computer Science and Information Engineering, Asia University, Taichung, Taiwan. Email: robertchh@gmail.com. 
\IEEEcompsocthanksitem Q. Wu is with the College of Computer, National University of Defense Technology, Changsha 410073, China. Email: wuqingbo@tj.kylinos.cn. 
\IEEEcompsocthanksitem R. Li is with Peng Cheng Laboratory, Shenzhen 518000, China. Email: lir@pcl.ac.cn.
\IEEEcompsocthanksitem AY. Zomaya is with the School of Computer Science, The University of Sydney, Sydney, Australia. Email: albert.zomaya@sydney.edu.au.
\IEEEcompsocthanksitem
Copyright (c) 20xx IEEE. Personal use of this material is permitted. However, permission to use this material for any other purposes must be obtained from the IEEE by sending a request to pubs-permissions@ieee.org.
}
 
 }
%
%

\markboth{}%
{Huang \MakeLowercase{\textit{\textit{et al.}}}:  Adaptive Processor Frequency Adjustment for Mobile Edge Computing with Intermittent Energy Supply}
%




\maketitle
\begin{abstract}
	With astonishing speed, bandwidth, and scale, Mobile Edge Computing (MEC) has played an increasingly important role in the next generation of connectivity and service delivery. Yet, along with the massive deployment of MEC servers, the ensuing energy issue is now on an increasingly urgent agenda. In the current context, the large-scale deployment of renewable-energy-supplied MEC servers is perhaps the most promising solution for the incoming energy issue.  Nonetheless, as a result of the intermittent nature of their power sources, these special design  MEC servers must be more cautious about their energy usage, in a bid to maintain their service sustainability as well as service standard. Targeting optimization on a single-server MEC scenario, we in this paper propose NAFA, an adaptive processor frequency adjustment solution, to enable an effective plan of the server's energy usage.  By learning from the historical data revealing request arrival and energy harvest pattern, the deep reinforcement learning-based solution is capable of making intelligent schedules on the server's processor frequency, so as to strike a good balance between service sustainability and service quality. The superior performance of NAFA is substantiated by real-data-based experiments, wherein NAFA demonstrates up to  20\% increase in average request acceptance ratio and up to 50\% reduction in average request processing time.
\end{abstract}
%
\IEEEpeerreviewmaketitle

\section{Introduction}
%
%
%
%

\subsection{Background and Motivations}
\IEEEPARstart{L}{ately}, Mobile Edge Computing (MEC) has emerged as a powerful computing paradigm for the future Internet of Things (IoTs) scenarios. MEC servers are mostly deployed in proximity to the users, with the merits of seamless coverage and extremely low communication latency to the users. Also, the MEC servers are light-weight, enabling their potential large-scale application in various scenarios. \par
Despite an attractive prospect, two crucial issues might be encountered by the real application: 
\begin{enumerate}
\item The large-scale deployment of grid-power MEC servers has almost exhausted the existing energy resource and resulted in an enormous carbon footprint. This in essence goes against the green computing initiative.
\item With millions or billions of small servers deployed amid every corner of the city, some of the locations could be quite unaccommodating for the construction of grid power facility, and even for those in a good condition, the construction and operation overhead of the power facility alone should not be taken lightly.
\end{enumerate}
\par
With these two challenges encountered during the large-scale application of MEC, we initiate an alternative usage of \textit{intermittent energy supply}. These supplies could be \textit{solar power}, \textit{wind power} or \textit{wireless power}, etc. Amid these alternatives, renewable energy, such as solar power and wind power, could elegantly address both the two concerns. The wireless power still breaches the green computing initiative, but can at least save the construction and operation cost of a complete grid power system for all the computing units.  \par
However, all of these intermittent energy supplies reveal a nature of unreliability: power has to be stored in a battery with limited capacity for future use, and this limited energy is clearly incapable of handling all the workloads when mass requests are submitted. With this concern, servers have two potential options to accommodate the increasing workloads:
\begin{enumerate}
\item They directly reject some of the requests (perhaps those require greater computation), in a bid to save energy for the subsequent requests.
\item They lower the processing frequency of their cores to save energy and accommodate the increasing workloads. Nevertheless, this way does not come without a cost: each request might suffer a prolonged processing time.
\end{enumerate}
\par
These heuristic ideas of energy conservation elicit the problem we are discussing in this paper. We are particularly interested in the potential request treatment (i.e., should we reject an incoming request, and if not, how should we schedule the processing frequency for it) and their consequent effects on the overall system performance. \par
The problem could become even more sophisticated if regarding the unknown and non-stationary request arrival and energy harvest pattern. Traditional rule-based methods clearly are incompetent in this volatile context: as a result of their inflexibility, even though they might work in a particular setting (a specific arrival pattern, for example), they may not work equally well if in a completely different environment. \par
Being motivated, we shall design a both effective and adaptive solution that is able to cope with the uncertainty brought by the energy supply and request pattern. Moreover, the solution should be highly programmable, allowing custom design based on the operators' expectations towards different performance metrics.   
\par

\subsection{Contributions }
The major contributions of our work are presented in the following:
\begin{enumerate}
	\item We have analyzed the real working procedure of an intermittent-energy-driven MEC system,  based on which, we propose an event-driven schedule scheme, which is deemed much matching with the working pattern of this system. 
	\item Moreover, we propose an energy reservation mechanism to accommodate the event-driven feature. This mechanism is novel and has not been available in other sources, to our best knowledge.
	\item Based on the proposed scheduling mechanism, we have formulated an optimization problem, which basically covers a few necessary system constraints and two main objectives: cumulative processing time and acceptance ratio.  
	\item We propose a deep reinforcement learning-based solution, dubbed {\color{black} \textbf{N}eural network-based \textbf{A}daptive \textbf{F}requency \textbf{A}djustment} (NAFA), to optimize the joint request treatment and frequency adjustment action.
	\item We do the experiments based on real solar data. By our experimental results, we substantiate the effectiveness and adaptiveness of our proposed solutions. In addition to the general results, we also develop a profound analysis of the working pattern of different solutions.
\end{enumerate}
\par
To the best knowledge of the authors, our main focus on event-driven scheduling for an intermittent energy-supplied system has not been presented in other sources. Also, our proposed solution happens to be the first trackable deep reinforcement learning solution to an SMDP model. It has the potential to be applied to other network systems that follow a similar event-driven working pattern. In this regard, we consider our novel work as a major contribution to the field. 
\section{Related Work}

\subsection{Edge intelligence}
{\color{black}
Artificial intelligence (AI) have been broadly studied thanks to their unlimited potential in image classification, natural language processing, and scheduling (e.g. \cite{huang2020ant}, \cite{wu2020collaborate}). To further boost the performance of AI, some literature has proposed MEC-combined solutions.  Khan et al. in \cite{khan2019deep} proposed a deep unified model for Face Recognition, and further integrated the model into the edge computing system. Experimental results show that such a combination significantly reduces data latency of processing the needed data and increases the real-time response.  Another interesting direction is to boost the performance of edge intelligent systems by a secure data-shared algorithm. To illustrate, Feng et al. in \cite{feng2020attribute} proposed an  Attribute-Based Encryption (ABE) parallel outsourced decryption for edge intelligent Internet of Vehicles (IoV). The parallel design increases decryption speed in the resource-constrained IoV context while maintaining the same security compared with the original ABE scheme. In another recent work \cite{yang2019efficient}, targetting the edge artificial intelligence scenario, Yang et al. proposed a multi pedestrian tracking method based on the rank constraint. Their proposed method is beneficial to eliminate ambiguous detection responses during association, and further increases the tracking performance in the intelligence edge device. 
\par
}
{\color{black}
\subsection{Reinforcement learning and its application in edge}
Reinforcement learning is another key technique that was frequently applied into the scheduling problems persisted in an edge-computing system. This particular technique usually bases upon two main genres of models, i.e., Markov Decision Process (MDP), and Semi Markov Decision Process (SMDP).
\subsubsection{MDP-based reinforcement learning}
\par
   Refs.\cite{chen_performance_2018,wei_dynamic_2019,min_learning-based_2019,xu2019optimized} are typical examples of edge Markov Decision Process (MDP) based reinforcement learning techniques that is applied into the scheduling problem in edge.} Specifically, in \cite{chen_performance_2018}, Chen \textit{et al.} combined the technique of Deep Q Network (DQN) with a multi-object computation offloading scenario. By maintaining a neural network inside its memory, the {\color{black}Mobile User} (MU) is enabled to intelligently select an offloading object among the accessible base stations.  Wei \textit{et al.} in \cite{wei_dynamic_2019} introduced a reinforcement learning algorithm to address the offloading problem in the IoT scenario, based on which, they further proposed a value function approximation method in an attempt to accelerate the learning speed. In \cite{min_learning-based_2019}, Min \textit{et al.} further considered the privacy factors in healthcare IoT offloading scenarios and proposed a reinforcement learning-based scheme for a privacy-ensured data offloading. 
\par
{\color{black}
The above works are all based on MDP, which means that their action scheduling is performed periodically based on a fixed time interval.
However, in the real scenario, the workflow of an agent (server or MU) is usually event-driven, and requires prompt scheduling response. A more reasonable setting is that the agent could promptly make the scheduling and perform actions once an event has occurred (e.g. a request arrives) but not wait until a fixed periodicity is met.}
\subsubsection{Event-Driven SMDP-based reinforcement learning }
\par
 Unlike the conventional MDP model, for an SMDP model, the time interval between two sequential actions does not necessarily need to be the same. As such, it is super matching for an event-triggered schedule, like the one we need for the requests scheduling.  \par
SMDP is a powerful modeling tool that has been applied to many fields, such as wireless networks, operation management, etc.  In \cite{zheng_smdp-based_2015}, Zheng \textit{et al.} first applied SMDP in the scenario of vehicular cloud computing systems. A discounted reward is adopted in their system model, based on which, the authors further proposed a value iteration method to derive the optimal policy. In \cite{baljon_smdp-based_2017}, SMDP is first applied to energy harvesting wireless networks. For problem-solving, the authors adopted a model-based policy iteration method. However, the model-based solution proposed in the above work can not address the problem when the state transition probability is unknown, and in addition, it cannot deal with the well-known curse-of-dimensionality issue. \par
To fix the gap, based on an SMDP model that is exclusively designed for a {\color{black} Narrow Band Internet of Things (NB-IOT)} Edge Computing System, Lei \textit{et al.} in \cite{lei_joint_2019} further proposed a reinforcement learning-based algorithm. However, the proposed algorithm is still too restricted and cannot be applied in our current studied problems, since several assumptions (e.g., exponential sojourn time between events) must be made in advance. Normally, these assumptions are inevitable for the derivation of state-value or policy-value estimation in an SMDP model, but unfortunately, could be the main culprit leading to great divergence between theory and reality. This potential drawback inspires us to \textbf{jettison all the assumptions} typically presented in an SMDP formulation in our research.\par
\subsection{DVFS and its optimization}
{\color{black}
 Dynamic Voltage and Frequency Scaling (DVFS) is a technique to adjust the frequency of a micro-processor, and is widely employed in cloud and edge to achieve energy conservation of computing. In the cloud, Wu et al. \cite{wu2019power} designed a scheduling algorithm for the cloud datacenter with the DVFS technique. Their proposed method efficiently assigns proper resources to jobs according to their requirements, and the authors claim that their approach, while efficiently reduces the energy consumption of a data center, does not sacrifice the system performance. In an edge environment, Dinh et al. in \cite{dinh2017offloading} jointly optimize the task allocation decision and its CPU frequency. By exploiting these benefits brought by proper allocation decisions and CPU frequency, the mobile device's energy consumption could be lower, while the task latency performance can be improved. The above works rely on frequency adjustment in the dimension of the task. Another study \cite{ajirlou2020machine2} has considered adaptive frequency adjustment at the instruction level. Essentially, they proposed to  classify individual instruction
into the corresponding propagation delay classes in real-time, and the clock frequency is accordingly adjusted to reduce the gap between the actual propagation delay and the clock period, and thereby reaping the benefits of dynamic frequency adjustment. The instruction-level optimization is also studied by an earlier work \cite{zhang2017bandits2}. Their main motivation is that the timing speculation (i.e., frequency) has a sweet spot for each instruction. Too much timing speculation (i.e., higher frequency, or too aggressive acceleration) could increase timing error probability, which may inversely hurt the performance. To find the sweet spot, a bandit algorithm is proposed to balance exploration and exploitation. 
\par In this work, we will focus on the task/job level optimization of the dynamic frequency, while leaving the instruction-level optimization a future work. 
}
\section{Problem Formulation}
\label{Problem Formulation}
%
%

\subsection{System Overview}
\begin{figure*}[!t]
\centering
\includegraphics[width=6.5in]{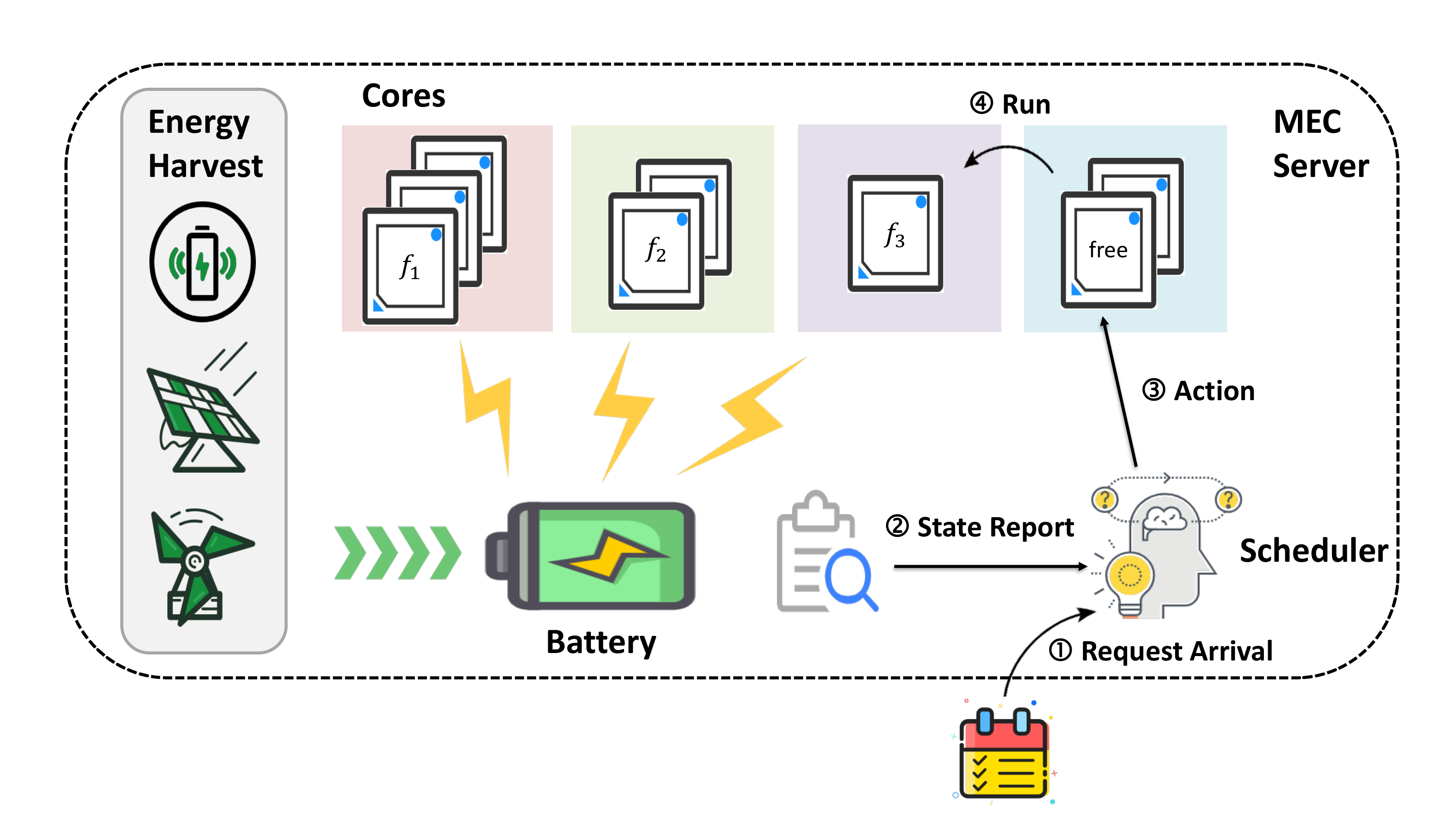}
\caption{System architecture for an intermittent-energy-supplied MEC system.}
\end{figure*}
In this paper, we target the optimization problem in a multi-users single-server MEC system that is driven by intermittent energy supply (e.g., renewable energy, wireless charging).  As depicted in Fig. 1, multiple central processing units (CPUs) and a limited-capacity battery, together with communication and energy harvesting modules, physically constitute an MEC server.  Our proposed MEC system makes scheduling on requests following an event-driven workflow, i.e., our request scheduling process is evoked immediately once a request is collected.  {\color{black} Our solution is different from the general batch-scheduling methods, which would require the requests to wait in a queue until the scheduler evokes (either periodically or until a sufficient number of requests are queued). Compared to batch-scheduling, event-driven scheduling could make a faster response on the incoming request, and therefore could ensure a higher quality of service. But we admit that since more information could be gathered when requests are scheduled in batch, better performance of batch scheduling may be gained in return.  } Formally, the proposed event-driven process can be specified by the following steps:
\begin{enumerate}
	\item Collect the request and check the current system status, e.g., battery and energy reservation status and CPU core status.
	\item Do schedule to the incoming request based on its characteristic and current system status. Explicitly, the scheduler is supposed to make the following decisions: 
	\begin{enumerate}
		\item Decide whether to accept the request or not.
		\item  If accepted, decide the \textit{processing frequency} of the request. We refer to \textit{processing frequency} as the frequency that a CPU core might turn to while processing this particular request. The scheduler should not choose the frequency that might potentially \textit{over-reserve} the currently available energy or \textit{over-load} the available cores. 
		\end {enumerate}
		\item Do energy reservation for the request based on the decided frequency. A CPU core cannot use more than the reserved energy to process this request.  
		\item The request is scheduled to the corresponding CPU core and start processing.
	\end{enumerate}\par
	Given that the system we study is \textbf{not} powered by reliable energy supply (e.g., coal power), a careful plan of the available energy is supposed to be made in order to promote the system performance.

\subsection{Formal Statement}
\begin{table}[!t]
	\caption{Key notations for problem formulation}
	\label{table notations}
	\centering
	\begin{tabular}{cc }
		\hline
	    Notations  & Meanings  \\
		\hline
		$a_i$ & action scheduled for the $i$-th request   \\
		$f_n$  &     $n$-th processing frequency option (GHz)       \\
		$d_i$  &     data size of the $i$-th request (bits)    \\
		$\nu$ & computation complexity per bit \\
		$\kappa$ & effective switched capacitance  \\
		$m$ & number of CPU cores \\
		$\tau_{a_i}$ & processing time of the $i$-th request \\
		$e_{a_i}$ & energy consumption of the $i$-th request \\
		$B_{i}$ & battery status when $i$-th request arrives \\
		$B_{max}$ & maximum battery capacity \\
		\multirow{2}{*}{$\lambda_{i,i+1}$/$\rho_{i,i+1}$} & captured/consumed energy between   \\ &  arrivals of the $i$-th and $(i+1)$-th request\\
		$S_{i}$ & reserved energy when $i$-th request arrives\\
		\multirow{2}{*}{$\Psi_i$} & number of working CPU cores \\ &when the $i$-th request arrives\\
		$\eta$ & tradeoff parameter\\
		\hline
	\end{tabular}
\end{table}

In this subsection, we shall formally introduce the optimization problem by rigorous mathematics formulation (key notations and their meanings are given in Table \ref{table notations}).  We consider the optimization problem for an MEC server with $m$ CPU cores, all of whose frequency can be adaptively adjusted via Dynamic Voltage and Frequency Scaling (DVFS) technique. Then we first specify the {\color{black}action} for the optimization problem.
\subsubsection{Action}
The decision (or action)  in this system specifies the treatment of the incoming request. Explicitly, it specifies 1) whether to accept the request or not,  2) the processing frequency of the request. Formally, let $i$ index an incoming request by its arrival order. The action for the $i$-th request is denoted by $a_i$. Explicitly, we note that:
\begin{itemize}
\item  $a_i \in \left\{0, 1,\dots,n \right \}$ denotes the action index of the CPU frequency at which the request is scheduled, where $n$ represents the maximum index (or total potential options) for frequency adjustment. For option $n$, $f_n$ GHz frequency will be scheduled to the request. 
\item Specially, when $a_i=0 $, the request will be rejected immediately. 
\end{itemize}
\subsubsection{ Processing Time and Energy Consumption }
The action we take might have a direct influence on the request \textit{processing time}, which can be specified by:
    \begin{align}
 \label{processing time}
\begin{split} 
\tau_{a_i}&=\begin{cases}    \frac{ \nu \cdot d_i   }{ f_{a_i}}  & a_i \in \{ 1, \dots,n  \} \\ 0 & a_i=0  \end{cases}
\end{split}
 \end{align}
 where we denote $d_i$ as the processing data size of an offloading request, $\nu$ as the required CPU cycles for computing one bit of the offloading data. Without loss of generality, we assume $d_i$, i.e., processing data size, as a stochastic variable while regarding $\nu$, i.e., required CPU cycles per bit, fixed for all requests.    \par
Also, by specifying the processing frequency, we can derive the \textit{energy consumption} for processing the request, which can be given as:
    \begin{align}
 \label{energy consumption}
\begin{split} 
e_{a_i}= \begin{cases} \kappa    f_{a_i}^2  \cdot  \nu \cdot d_i      &   a_i \in \{ 1,\dots,n  \}   \\ 0 & a_i=0\end{cases}
\end{split}
 \end{align}
where we denote $\kappa$ as the effective switched capacitance of the CPUs.

\subsubsection{Battery Status}
Recall that the MEC server is driven by intermittent energy supply, which indicates that the server has to store the captured energy in its battery for future use. Concretely, we shall introduce a virtual queue, which serves as a measurement of the server's current battery status. Formally, we let a virtual queue, whose backlog is denoted by $B_i$, to capture the energy status when the $i$-th request arrives. $B_i$ evolves following this rule:
\begin{equation}
\label{battery status}
 B_{i+1}=\min \left \{B_{i}+ \lambda_{i,i+1}- \rho_{i,i+1}  ,  B_{max} \right\}
\end{equation}
Here,
\begin{enumerate}
\item  $\lambda_{i,i+1}$ represents the amount of energy that was captured by the server between the arrivals of the $i$-th and $(i+1)$-th request. 
\item $\rho_{i,i+1}$ is the consumed energy during the same interval. It is notable that the consumed energy i.e., $\rho_{i,i+1}$, is highly related to the frequency of the current running cores, due to which, is also relevant to the past \textit{actions}, regarding the fact that it is the past actions that determine the frequency of the current running cores.
\item  $B_{max}$ is the maximum capacity of the battery. we use this value to cap the battery status since the maximum energy status could not exceed the full capacity of the server's battery. 
\end{enumerate}
\subsubsection{Energy Reservation Status}
Upon receiving each of the coming requests, we shall reserve the corresponding amount of energy for which so that the server could have enough energy to finish this request. This reservation mechanism is essential in maintaining the stability of our proposed system. \par 
To be specific, we first have to construct a virtual queue to record the \textit{energy reservation status}. The energy reservation queue, with backlog $S_i$, evolves between request arrivals following this rule:  
\begin{equation}
\label{energy reservation status}
 S_{i+1}= \max \left \{S_{i}+ e_{a_i} - \rho_{i,i+1}   ,0 \right \}
\end{equation}
where $\rho_{i,i+1}$ is the consumed energy (same definition in Eq. (\ref{battery status})) and  $e_{a_i}$ is the energy consumption for the $i$-th request (same definition in Eq. (\ref{energy consumption})).  By this means, $S_{i+1}$ exactly measures how much energy has been reserved by the {\color{black} first} to the $i$-th request \footnote{Some may wonder the rationale behind the subtraction of $\rho_{i,i+1}$. Considering the case that in a specific timestamp (e.g., the timestamp that the $i$-th request arrives), the processing of a past request (e.g., the $(i-1)$-th request) could be unfinished, but have already consumed some of the reservation energy. Then, the reservation energy for it should subtract the already consumed part.}.
\par
\subsubsection{Energy Constraint}
By maintaining the \textit{energy reservation status}, we can specify \textit{energy constraint} to control the action when the energy has already been full-reserved. The constraint can be given as follows:
\begin{equation}
 S_{i}+ e_{a_i} \leq B_{i} 
\end{equation}
By this constraint, we ensure that the request has to be rejected if not enough energy (that has not been reserved) is available.

\subsubsection{Resource Constraint}
Recall that we assume a total number of $m$ CPU cores are available for request processing. When the computation resources have already been full-loaded, the system has no option but to reject the arrived requests. So, we introduce the following constraint:
\begin{equation}
\label{resource constraint}
\Psi_i+ \mathbb{I}\{a_i \neq 0\} \leq m
\end{equation}
where $\Psi_i$ denotes the number of currently working CPU cores. By this constraint, we ensure that the server could not be overloaded by the accepted request.
\subsubsection{Optimization Problem}
Now we shall formally introduce the problem that we aim to optimize, which is given as follows:
 \begin{equation}
 \begin{split}
 \textit{ (P1)}  &  \quad \textbf{Obj1:} \quad \min_{\{a_i\}}  \sum_{i=1}^{\infty}  \tau_{a_i}  \quad  \\
&  \quad \textbf{Obj2:} \quad  \max_{\{a_i\}}    \sum_{i=1}^{\infty} \mathbb{I}\{a_i \neq 0\}  \\
  &\quad\textbf{C1:} \quad  S_{i}+ e_{a_i} \leq B_{i}   \\ 
   & \quad \textbf{C2:}    \quad \Psi_i+ \mathbb{I}\{a_i \neq 0\} \leq m
  \end{split}
 \end{equation} \par
There are two objectives that we need to consider in this problem: 1) the first objective is the cumulative \textit{processing time} of the system, 2) and the second objective is the cumulative \textit{acceptance}. Besides, two constraints, i.e., \textit{Energy Constraint} and \textit{Resource Constraint}, have been covered in our system constraints. \par 
Obviously,  \textit{P1} is unsolvable due to the following facts: 
\begin{itemize}
	\item \textbf{Existence of stochastic variables.} Stochastic variables, e.g., $d_{i}$ (data size of requests),   $\lambda_{i,i+1}$ (energy supplied) and request arrival rate, persist in \textit{P1}. 
	\item \textbf{Multi-objective quantification.} The two objectives considered in our problem are mutually exclusive and there is not an explicit quantification between them.
\end{itemize}
To bridge the gap, we 1) set up a tradeoff parameter to balance the two objectives and transform the problem into a single objective optimization problem, 2) transform the objective function into an expected form. This leads to our newly formulated \textit{P2}:   
 \begin{equation}
 \begin{split}
\label{P2}
 \textit{ (P2)}  &\quad  \quad  \max_{\{a_i\}}    \sum_{i=1}^{\infty} \mathbb{E} \left[ \mathbb{I}\{a_i \neq 0\}-\eta \tau_{a_i}  \right] \\
  &\quad\textbf{C1:} \quad  S_{i}+ e_{a_i} \leq B_{i}   \\ 
   & \quad \textbf{C2:}    \quad \Psi_i+ \mathbb{I}\{a_i \neq 0\} \leq m
  \end{split}
 \end{equation} 
where $\eta$ serves as the tradeoff parameter to balance cumulative \textit{acceptance} and cumulative \textit{ processing time}.  \textit{P2} is more concrete  after the transformation, but still, we encounter the following challenges when solving \textit{P2}:
\begin{itemize}
	\item \textbf{Exact constraints.} Both C1 and C2 are exact constraints that should be strictly restricted for each request. This completely precludes the possibility of applying an offline deterministic policy to solve the problem, noticing that $B_i$, $S_i$ in C1 and $\Phi_i$ in C2 are all stochastic for each request.
	\item \textbf{Unknown and non-stationary supplied pattern of energy.} The supplied pattern of intermittent energy is highly discrepant, varying from place to place and hour to hour\footnote{Considering the energy harvest pattern of solar power. We have a significant difference in harvest magnitude between day and night.}. As such, the stochastic process $\lambda_{i,i+1}$ in Eq. (\ref{battery status}) may not be stationary, i.e., $\lambda_{i,i+1}$ samples from an ever-changing stochastic distribution that is relevant with request order $i$, and moreover, it is unknown to the scheduler. 
	\item \textbf{Unknown and non-stationary arrival pattern.}  The arrival pattern of the request's data size is unknown to the system, and also, could be highly discrepant at temporal scales.
	\item\textbf{Unknown stochastic distribution of requests' data size.}  The processing data size could vary between requests and its distribution is typically unknown to the scheduler.
	\item \textbf{Coupling effects of actions between requests.} The past actions towards an earlier request might have a direct effect on the later scheduling process (see the evolvement of $S_{i}$ and $B_i$)
\end{itemize}
Regarding the above challenges, we have to resort to an online optimization solution for problem-solving. The solution is expected to learn the stochastic pattern from the historical knowledge, and meanwhile, can strictly comply with the system constraint. 
\par
{\color{black}
\subsection{Connection Between Our Problem and Knapsack}
Our defined problem can be viewed in a classical knapsack view. The key problem of the knapsack is that: if we have $N$ items (indexed by $i$) with a value $v_i$ and a weight $w_i$, which subset of items should we put into a knapsack with capacity $B$, such that the total value of collected items maximized? If we view a request as an item,  the energy consumption of processing it as weight, the reward of accepting it as value, and the available energy as knapsack capacity, our problem can be viewed as an extension of the knapsack problem. But the key difference is that:
\begin{itemize}
	\item The item list is not finite and fixed. The item (request) arrives across the timeline with the unknown pattern of weights (energy consumption) and values (rewards). Before a request’s arrival, the scheduler can not access the request's value and weight. In addition, the scheduler is supposed to make the decision once a request arrives.
	\item The knapsack capacity (available energy) is growing across the timeline since the server continuously harvests energy. Also, how much energy will be harvested in the future is unknown to the scheduler.
	\item There is another constraint other than the available energy constraint. The number of requests that can be simultaneously hosted by the server is constrained (See Eq. (6)). 
	\item The action space is no longer 0-1, as in the original knapsack problem. The action space of our problem includes whether to accept the request, as well as the processing frequency of it. 
\end{itemize}
The main deviation between our problem and knapsack is caused by the event-driven scheduling. The scheduler should make a prompt decision once an item (request) arrives but is not allowed to make the decision after receiving a batch of requests. In terms of the batch-scheduling case, if the batch size is sufficiently large, the problem can be reduced to a knapsack problem with finite action set and multiple knapsack capacity (i.e., multiple constraints). However, when using event-driven scheduling (do scheduling once a request arrives), this reduction may not hold.
}
\section {Deep Reinforcement Learning-Based Solution}
\subsection{Formulation of a Semi-Markov Decision Process}
To develop our reinforcement learning solution, we shall first transform the problem into an \textit{Semi-Markov Decision Process} (SMDP) formulation. In our SMDP formulation, we assume the system \textit{state} as the system running status when a request has come, or when the scheduler is supposed to take \textit{action}. After an action being taken by the scheduler, a \textit{reward} would be achieved, and the state (or system status) correspondingly transfers. It follows an \textit{event-driven} process, that is, the scheduler decides the action once a request has come but does nothing while awaiting. And as a result, the time interval between two sequential actions may not be the same. This is the core characteristic of an SMDP model and is the major difference from a normal MDP model.  \par
Now we shall specify the three principal elements (i.e., states, action, and rewards) in our SMDP formulation in sequence, and please note that most of the notations we used henceforth are consistent with those in Section \ref{Problem Formulation}.  
\subsubsection{System States}
A system state is given as a tuple: 
\begin{equation}
\label{state formulation}
s_i \triangleq \{ T_i, B_i, S_i, \psi_{i,1} \dots, \psi_{i,n}, d_i \} 
\end{equation}
Explicitly, 
\begin{itemize}
\item $T_i$ is the \textbf{24-hour-scale local time} when the $i-$th request arrives in the MEC server. We incorporate this element in our state in order to accommodate the temporal factor that persists in the energy and request pattern. 
\item $B_i$ is the \textbf{battery status}, same as we specify in Eq. (\ref{battery status}).
\item $S_i$ is the \textbf{energy reservation status}, same as we specify in Eq. (\ref{energy reservation status}).
\item $\psi_{i,n}$ denotes the\textbf{ running CPU cores in the frequency of $f_n$ GHz}. Meanwhile, it is intuitive to see that $\Psi_i= \sum_{n^{\prime}=1}^n\psi_{i,n}$ where $\Psi_i$ is the total running cores we specify in Eq. (\ref{resource constraint}).
\item $d_i$ is the \textbf{data size} of the $i$-th request. 
\end{itemize}
Information captured by a state could be comprehended as the current system status that might support the action scheduling of the incoming $i$-th request. More explicitly, we argue that the formulated states should at least cover the system status that helps construct a \textit{possible action set}, but could cover more types of informative knowledge to support the decision (e.g., $T_i$ in our current formulation, which will be formally analyzed later). The concept of \textit{possible action set} would be given in our explanation of \textit{system actions}.
\subsubsection{System Actions}
\label{action set}
Once a request arrives, given the current state (i.e., the observed system status),  the scheduler is supposed to take action, deciding the treatment of the request. Indicated by the system constraints C1 and C2 in \textit{P2}, we are supposed to make a restriction on the to-be taken action. By specifying the states (or observing the system status), it is not difficult to find that we can indeed get a closed-form possible action set, as follows:
\begin{equation}
 a_i	\in \mathcal{A}_{s_i} \triangleq \{ a | \Psi_i+ \mathbb{I}\{a \neq 0\} \leq m,   S_{i}+ e_{a} \leq B_{i}  \}  
\end{equation}
where $\Psi_i$, $B_{i}$ and $S_i$ are all covered in our state formulation. By taken action from the defined possible action set, we address challenge 1) \textbf{exact constraints},  that we specify  in \textit{P2}.

\subsubsection{System Rewards}
Given state and action, a system reward will be incurred, in the following form:
 \begin{equation}
 \label{reward}
r(s_i,a_i)= \mathbb{I}\{a_i \neq 0\}- \eta \tau_{a_i} 
\end{equation}
The reward of different treatments of a request is consistent with our objective formulation in \textit{P2} (see Eq. (\ref{P2})). The goal of our MDP formulation is to maximize the expected achieved rewards, which means that we aim to maximize $ \sum_{i=1}^{\infty} \mathbb{E} \left[ \mathbb{I}\{a_i \neq 0\}- \eta \tau_{a_i}  \right]$, the same form with the objective in \textit{P2}. Besides, here we can informally regard $\mathbb{I}\{a_i \neq 0\}$ as a "real reward" of accepting a request and $\tau_{a_i}$ as the "penalty" of processing a request. 
\subsubsection{State Transferred Probability}
After an action being taken, the state will be transferred to another one when the next arrival of the request occurs. The state transferred probability to a specific state is assumed to be \textit{stationary} given the current state, as well as the current action, i.e., we need to assume that:
\begin{equation}
	\label{state transfer probability }
p(s_{i+1} |s_i,a_i)= \text{a constant}
\end{equation}
\par
This assumption is the core part in our MDP formulation, which is our main motivation to cover $T_i$ and $\psi_{i,1} \dots, \psi_{i,n}$ (rather than $\Psi_{i}$) in our state formulation. {\color{black}
	By this constant state transferred probability assumption, we are allowed to estimate the transferred probability, and thereby given a fighting chance to maximize the achieved rewards in an online fashion. To justify this assumption given our defined state formulation, we refer the readers to Appendix \ref{appendix A} for a detailed explanation. {
		\color{black}
		And to make the MDP formulation easier to grasp, we prepare a scheduling example, which has been moved to Appendix \ref{appendix B} due to the space limit. }
} \par
\par
\subsection{State-Action Value Function}
\label{old State-Action Value Function}
Recall that our ultimate goal is to maximize the expected cumulative reward. For this purpose, we leverage Bellman optimality to find the optimal discounted state-action value, i.e., $Q_{\pi^*}(s,a)$, as follows:
\begin{equation}
	\begin{split}
		\label{final iterate update function}
		&Q_{\pi^*}(s,a) =  \mathbb{E}_{s^{\prime}}  \left[   r(s,a)+  \beta  \max_{a^{\prime} \in \mathcal{A}_{s^{\prime}}}  Q_{\pi^*}(s^{\prime},a^{\prime})      \right]
	\end{split}
\end{equation} 
where $\pi^*$ is the optimal policy under discounted formulation, and $\beta$ is the discounted factor. {\color{black} Detailed explanation about how to derive $Q_{\pi^*}(s,a)$ is available in Appendix \ref{Appendix C}.} This state-action value., i.e., $Q_{\pi^*}(s,a)$ indeed specifies the value of taking a specific action, which is basically composed of two parts:
\begin{enumerate}
\item The first part is the expected reward that is immediately obtained after actions have been taken, embodied by $\mathbb{E}[r(s,a)]$.
\item The second part is the discounted future expected rewards after the action has been taken, embodied by $ \mathbb{E}[\beta  \max_{a^{\prime} \in \mathcal{A}_{s^{\prime}}}  Q_{\pi^*}(s^{\prime},a^{\prime})] $. This literally addresses challenge 5) \textbf{Coupling effects of actions between requests} we proposed below \textit{P2}. The coupling effects are in fact embodied by these discounted expected rewards: by taking a different action, the state might experience a different transfer probability and thereby making the future rewards being affected. By this MDP formulation, we are enabled to have a concrete model of this coupling effect to the future. 
\end{enumerate}\par
Besides, specified by the proposed challenges below \textit{P2}, we know that both the energy and request arrival pattern, as well as the data size distribution, are all unknown to the scheduler, which means that it is hopeless to derive the closed-form $Q_{\pi^*}(s,a)$. More explicitly, since the state transferred probability $p(s^{\prime} |s,a)$ (or $p(s_{i+1} |s_i,a_i)$ in Eq. (\ref{state transfer probability })) is an unknown constant, we cannot expand the expectation in Eq. (\ref{final iterate update function}), making $Q_{\pi^*}(s,a)$ unachievable in an analytical way. Without knowledge of $Q_{\pi^*}(s,a)$, we are unable to fulfill our ultimate goal, i.e., to derive $\pi^*$. \par
Fortunately, there still exists an alternative path to derive  $Q_{\pi^*}(s,a)$. (Henceforth, we use $Q(s,a)$ to denote $Q_{\pi^*}(s,a)$ for sake of brevity). If we have sufficient amount of  data over a specific state and action (captured by a set $X_{s,a}$), each piece of which shapes like this 4-element tuple: $x \triangleq (s, a, r(s,a) , s^{\prime})$, then we can optimize $Q(s,a)$ by minimizing the \textit{estimation error}, namely:
\begin{equation}
\begin{split}
\label{estimate Q2}
\frac{1}{|X|} \sum_{x \in X} \left[ Q(s,a)- \left(   r(s,a)+  \beta  \max_{a^{\prime} \in \mathcal{A}_{s^{\prime}}}  Q(s^{\prime},a^{\prime})  \right)     \right]^2
\end{split}
\end{equation}
where $X=X_{s_1,a_1} \cup \dots \cup X_{s_1,a_n} \cup \dots$ captures all the available data.
In this way, we do not need prior knowledge about the transferred probability i.e., $p(s^{\prime} |s,a)$, but we learn it from the real state transferred data. This learning-based method completely resolves the \textbf{unknown distribution issues} we raised below \textit{P2}. \par
However, due to \textbf{infinite amount of states} in our formulated problem, the problem is still intractable.  We simply cannot iteratively achieve $Q(s,a)$ for every possible state-action pair, but at least, this initial idea points out a concrete direction that elicits our later double deep Q network solution.
\subsection{A Double Deep Q Network Solution}
In the previous section, we provide an initial idea about how to derive  $Q(s,a)$ in a learning way. But we simply cannot record $Q(s,a)$ for each state due to the unbounded state space. This hidden issue motivates us to use a neural network to predict the optimal state-action value (abbreviated as \textit{Q value} henceforth). Explicitly, we input a specific state to the neural network and output the Q value for action $1$ to $n$. By this means, we do not need to record the Q value, but it is estimated based on the output after going through the neural network. \par
 Formally, the estimated Q value can be denoted by $Q\left(s, a ; \theta_{i}\right)$ where $\theta_{i}$ denotes the parameters of the neural network (termed Q network henceforth) after $i$ steps of training. And following the same idea we proposed before, we expect to minimize the estimation error (or \textit{loss} henceforth), in this form:
\begin{equation}
\begin{split}
\label{Loss}
&L_{i}\left(\theta_{i};\tilde{X}_{i}\right)\\
=&\frac{1}{|\tilde{X}_i|} \sum_{x \in \tilde{X}_i} \left[ Q(s,a; \theta_{i})- \left(   r(s,a)+  \beta  \max_{a^{\prime} \in \mathcal{A}_{s^{\prime}}}  Q(s^{\prime},a^{\prime}; \theta_{i})  \right)     \right]^2
\end{split}
\end{equation}
{\color{black}
where,
\begin{itemize}
	\item $x \triangleq (s, a, r(s,a) , s^{\prime})$ is a tuple that captures a piece of data (which includes state, action, reward and next states). 
	\item  $\tilde{X}_i$ is a sub-set of total available data when doing training for the $i$-th step, which is often referred to as \textit{mini-batch}. We introduce such a concept here since the data acquiring process is an online process. In other words, we do not have all the training data naturally, but we obtain it through continuous interaction (i.e., acting action) and continuous policy update. As a result of this continuous update of data, it is simply not "cost-effective" to involve all the historical data we currently have in every step of training. As a refinement, only a subset of the data, i.e., $\tilde{X}_i$, is involved in the loss back-propagating for each step of training.
	\item $|\tilde{X}_i|$ is the cardinality of mini-batch $\tilde{X}_i$.
	\item $\theta_{i} \in \mathbb{R}^d$ denotes the parameters of the Q network. 
	\item  $Q(s,a; \theta_{i})$ is the predicted Q value for action $a$ via feeding state $s$ to the Q network.  
	\item $L_{i}\left(\theta_{i};\tilde{X}_{i}\right)$ is the loss function that we need to minimize by adjusting $\theta_{i}$ under mini-batch  $\tilde{X}_i$. 
	\item $\beta$ is the discounted factor.
\end{itemize}
}
 \par
\begin{figure*}[!hbtp] 
	\begin{equation}
		\begin{aligned}
			\label{double q Loss}
			&L_{i}\left(\theta_{i};\tilde{X}_{i}\right)=\frac{1}{|\tilde{X}_i|} \sum_{x \in \tilde{X}_i} \left[Q\left(s, a ; \theta_{i}\right)- \left(r(s,a) + \beta  Q\left(s^{\prime},  \mathop{\arg\max} _{a^{\prime} \in \mathcal{A}_{s^{\prime}}}Q(s^{\prime},a^{\prime}; \theta_{i}); \theta_{i}^{-}\right)    \right )\right]^{2}
		\end{aligned}
	\end{equation}
	{\color{black}where $\theta_{i} \in \mathbb{R}^d$ and  $\theta_{i}^{-} \in \mathbb{R}^d$ respectively denote the parameters of original Q network and target nework.  $Q(s,a; \theta_{i})$ is the predicted Q value for action $a$ via feeding state $s$ to the \textbf{target} Q network with network parameters $\theta_{i}^{-}$.   Other notations are consistent with those in Equation (\ref{Loss}).}
\end{figure*}
{\color{black}
However, the above-defined loss, though is quite intuitive, would possibly lead to training instability of the neural network, since it is updated too radically (see \cite{mnih_human-level_2015}). As per \cite{van2016deep}, a double network solution would ensure a more stable performance.  In this solution, we introduce another network, known as \textit{target network}, whose parameters are denoted by $\theta_{i}^{-}$. The target network has exactly the same network architecture as the Q network, and its parameters would be overridden periodically by the Q network. Informally, it serves as a "mirror" of the past Q network, which significantly reduces the training variation of the current Q network. The re-written loss function after adopting the double network architecture can be viewed in Eq. (\ref{double q Loss}) (located at the top of this page).
}
\par
\begin{algorithm}[h]  
        \caption {Training Stage of NAFA} 
        \begin{algorithmic}[1] 
         \REQUIRE ~~\\
         Initial/minimum exploration factor, $\epsilon_0$/ $\epsilon_{min}$;\\
	Discount factor for rewards/exploration factor, $\beta$/ $\xi$;\\
	
       	Tradeoff Parameter,  $\eta$; Learning rate $\gamma$;\\ 
	Update periodicity of target nework,  $\zeta$;\\
	Steps per episode, $N_{max}$; Training episodes, $ep_{max}$;\\
	Batch size, $|\tilde{X}_i|$; Memory size, $X_{max}$;\\
        \ENSURE~~\\
        After-trained network parameters; $\theta_{final}$
	\STATE Initialize $\theta_i$ and $\theta_{i}^{-}$  with arbitrary values
     	\STATE Initialize $i=1$, $\epsilon=\epsilon_0$
	\STATE Initialize empty replay memory $X$
	\STATE Initialize enviroment (or system status) 
	\FOR{$ep \in \{1,2, \dots,ep_{max}\}$}
     	\STATE Wait until the first request comes
     	\REPEAT
      \STATE Observe current system status $s$
     	\IF{ $\text{random()}<\epsilon$}
     	\STATE Randomly select action $a$ from $\mathcal{A}_s$
     	\ELSE
     	\STATE $
a=\operatorname{argmax}_{a \in \mathcal{A}_s } Q\left(s, a ; \theta_i\right)$
     	\ENDIF
     \STATE	Perform action $a$ and realize reward $r(s,a)$
	\STATE     Wait until the next request comes
	\STATE   Observe current system status $s^{\prime}$
     \STATE	Store $(s,a,r(s,a),s^{\prime})$ into replay memory $X$
     \STATE       Sample minibatch $\tilde{X}_{i} \sim X$
     \STATE Update parameter $\theta_{i+1}$  following  \begin{align} \label{gradient descent}\theta_{i+1}= \theta_{i}-\gamma \nabla_{\theta_{i}} L_i\left(\theta_{i};\tilde{X}_{i} \right)\end{align}
     \STATE $\epsilon_i=\epsilon_{min} + (\epsilon_{0}-\epsilon_{min}) \cdot \exp(-i /\xi )$
       \IF{ $i \mod \zeta=0$}
     \STATE  $\theta_{i+1}^{-}=\theta_{i+1}$ 
     \ELSE 
     \STATE  $\theta_{i+1}^{-}=\theta_{i}^{-}$ 
     \ENDIF
     \STATE $i=i+1$
           \UNTIL{ $i>ep\cdot N_{max}$ }
     \STATE Reset the environment (or system status)
     \ENDFOR
     \STATE  $\theta_{final}=\theta_{i}$
        \end{algorithmic} 	
        \label{Training Algorithm}	
	\end{algorithm}
Now we shall formally introduce our proposed solution termed \textbf{N}eural network-based \textbf{A}daptive \textbf{F}requency \textbf{A}djustment (NAFA), whose running procedure on training and application stage are respectively shown in Algorithm \ref{Training Algorithm} and Algorithm \ref{Application Algorithm}. Overall, the running procedure of NAFA can be summarized as follows:
\begin{enumerate}
	\item \textbf{Initialization:} NAFA first initializes the Q network with arbitrary values and set the exploration factor to a pre-set value. 
	\item \textbf{Iterated Training:} We divide the training into several episodes of training. The environment (i.e., the system status) will be reset to the initial stage each time an episode ends. In each episode, after the first request comes, the following sub-procedures perform in sequence:
	\begin{enumerate}
		
		\item \textbf{Schedule Action:}  NAFA observes the current system status $s$ (or state in our MDP formulation). Targeting the to-be scheduled request (i.e., the $i$-th request), NAFA adopts a $\epsilon$-greedy strategy. Explicitly, with probability $\epsilon$, NAFA randomly explores the action space and randomly selects an action. With probability $1-\epsilon$, NAFA greedily selects action based on the current system status $s$ and the current Q network's output.
		\item \textbf{Interaction:} NAFA performs the decided action $a$ within the \textbf{environment}. A corresponding reward would be realized as per Eq. (\ref{reward}). Note that the \textbf{environment} here could be real \textbf{operation enviroment} of a MEC server, or could be a \textbf{simulation environment} that is set up for a training purpose.
		\item \textbf{Data Integration:} After the interaction, NAFA sleeps until a new request has arrived. Once evoked, NAFA checks the current status $s^{\prime}$ (regarded as next state in the current iteration) and stores it into \textit{replay memory} collectively with current state $s$, action $a$, and realized reward $r(s,a)$. The replay memory has a maximum size $X_{max}$. If the replay memory goes full, the oldest data will be replaced by the new one.
		\item \textbf{Update Q Network:} A fixed batch size of data is sampled from replay memory to the minibatch $\tilde{X}_i$. Then, NAFA performs back-propagation with learning rate $\gamma$ on the Q network using samples within minibatch $\tilde{X}_i$ and the loss function defined in Eq. (\ref{double q Loss}).
		\item \textbf{Update Exploration Factor and Target Q Network:} NAFA discounts $\epsilon$ using a discount factor $\xi$ and update the target network at a periodicity of $\zeta$ steps. After that, NAFA starts a new training iteration.
\end{enumerate}
\begin{algorithm}[h]  
	\caption {Application Stage of NAFA} 
	\begin{algorithmic}[1] 
		\REQUIRE ~~\\
		After-trained network parameters; $\theta_{final}$
		\ENSURE~~\\
		Action sequence; $a_1, a_2, \dots$
		\STATE $i=1$
		\FOR{each request comes}
		\STATE Observe system status $s_i$
		\STATE $a_i=\operatorname{argmax}_{a \in \mathcal{A}_s} Q\left(s_i, a; \theta_{final}\right)$
		\STATE Perform action $a_i$
		\STATE $ i=i+1$
		\ENDFOR
	\end{algorithmic} 	
	\label{Application Algorithm}	
\end{algorithm}
\item \textbf{Application:} After training has finished (i.e., it has gone through $ep_{max} \cdot N_{max}$ steps of training), the algorithm uses the obtained Q network (whose parameters denoted by $\theta_{final}$) to schedule requests in the real \textbf{operation enviroment}, i.e., based on current state $s_i$,  action $a_i=\operatorname{argmax}_{a \in \mathcal{A}_{s_i}} Q\left(s_i, a; \theta_{final}\right)$ will be taken for each request. 
\end{enumerate}
\subsection{Discussion}
{\color{black}
\subsubsection{Computation complexity of NAFA}
Suppose we use a neural network which contains $n$ linear hidden layers, $m$ training samples, and $j$ nodes in each layer as the training model of NAFA. For forward propagating from a layer to its next layer, we have $\mathcal{O}(m*j^2)$ computation complexity, since the weights matrix between connected layers is in the demension of $j^2$, and we need to multiply the inputs (i.e., $m$ training samples) with the weights. After the matrix multiplcation, we apply the activation function to the output of each layer, which has the computation complexity of $\mathcal{O}(m*j)$. So, the forwarding complexity of NAFA for each layer is $\mathcal{O}(m*j+m*j^2 )$, and for $n$ layers in total it is $\mathcal{O}(n*(m*j+m*j^2) ) =\mathcal{O}(n*m*j^2)$. For the backward process, we compute the error signals for each node, and backward it to the previous layers. The time complexity is thus in the same scale with forwarding, which is also $\mathcal{O}(n*m*j^2)$. As such, the total computation complexity of NAFA's training process is  $\mathcal{O}(n*m*j^2)$.
\subsubsection{The way NAFA addresses the five challenges proposed below P2 }
\begin{itemize}
		\item \textbf{Exact constraints.} NAFA defines an action set based on the current state, so NAFA will not take invalid actions that may possibly violate the exact constraint (See \ref{action set}). 
	\item \textbf{Unknown and non-stationary supplied pattern of energy.} NAFA incorporates the local time into its state formulation to make the state transformation stationary (See Observation \ref{stationary energy arrival} in Appendix \ref{appendix A}), and uses the historical data to predict the value of a state (See our loss function in Eq.(\ref{double q Loss}) on how we predict the value), but remains agnostic about the real supplied pattern of energy.
	\item \textbf{Unknown and non-stationary arrival pattern.}  As the arrival pattern is also related to time, the local time incorporated into NAFA's states could make the state transformation stationary (See Observation \ref{stationary request arrival} in Appendix \ref{appendix A}). Also, NAFA is agnostic to the real request arrival pattern, but use the learning techniques to support decision.
	\item\textbf{Unknown stochastic distribution of requests' data size.}  We incorporate the data size of an incoming reqeust into NAFA's state formulation, based on which, NAFA is enabled to estimate the reward of computing the incoming request. Again, by our learning design, NAFA does not need to know the data size distribution, but can only care about the value of a specific state, which is learned by the historical data.  
	\item \textbf{Coupling effects of actions between requests.} To address this challenge, NAFA incorporates two parts when estimating the value of an state-action pair (aka Q value, see section \ref{old State-Action Value Function} ). The first part is the immediate reward of accepting a request, and the second part is the future rewards that is closely related to the state transformation. The coupling effects of action is profiled in the futures rewards, that is, taking an action may impact the state transformation, and thereby affecting the future reward after this action. Our state-value formulation accounts for this coupling effects of action. 
\end{itemize}
\subsubsection{Notes for implementation of NAFA}
Recall that NAFA is required to estimate the processing time and energy consumption of a request using Eq. (\ref{processing time}) and Eq. (\ref{energy consumption}). However, we state that such a calculation is replaceable. One can take advantage of other formulation to calculate these two metrics if her formulations are  closer to the real energy consumption and processing time of a request. Also, we need to clarify that these estimations do not necessarily need to be 100\% exact to the real quantities. For energy consumption, we allow the estimation to be an {\color{black} over-estimation}. By doing so, the request still would not fail in the middle of computation, since we have reserved more energy based on the overestimated expected energy consumption \footnote{\color{black}We note that when applying overestimation of energy consumption, we need to slightly modify Eq. (\ref{energy reservation status}) in our framework in order to maintain the normal operation. Once a request finishes, the server should check how much energy is really used by processing the requests, and derive how much energy has been over-reserved by the previous reservation. Then this quantity should be subtracted to eliminate the estimation error. }. For the processing time, which only affects the reward calculation, should not affect the working mechanism of the whole system, but of course, an accurate estimation certainly help NAFA to make a properer decision. 
}	

\section{Experiment}
\subsection{Setup}
\subsubsection{Basic setting}
\begin{itemize}
\item \textbf{Programming and running environment:}
We have implemented NAFA\footnote{Source codes of NAFA and baseline methods, as well as the simulation environment and dataset, are all available in https://github.com/huangtiansheng/NAFA} and the simulation environment on the basis of PyTorch and gym (a site package that is commonly used for environment construction in RL). Besides, all the computation in our simulation is run by a high-performance workstation (Dell PowerEdge T630 with 2xGTX 1080Ti).
\item \textbf{Simulation of energy arrival:}
In our simulation, NAFA is deployed on an MEC server which is driven by solar power (a typical example of intermittent energy supply). To simulate the energy arrival pattern, we use the data from HelioClim-3\footnote{http://www.soda-pro.com/web-services/radiation/helioclim-3-archives-for-free}, a satellite-derived solar radiation database. Explicitly, we derive the Global Horizontal Irradiance (GHI) data in Kampala, Uganda (latitude 0.329, longitude 32.499) from 2005-01-01 to 2005-12-31 (for train dataset) and 2006-01-01 to 2006-12-31 (for test dataset). Note that the GHI data in some days are missing in the dataset, so we use 300 out of 365 days of intact data respectively from the train and test dataset during our experiment. By the GHI data \footnote{\color{black} The GHI data in Helioclim is discrete by hours, so the time unit mentioned henceforth is 1 hour.  }, we calculate the amount of arrival energy during a given time interval $[t_1,t_2]$ ({\color{black}where $t_1$ and $t_2$ are integers}), as the following form: 
{\color{black}
\begin{equation}
	\lambda_{t_1,t_2}= \sum_{t=t_1}^{t_2}\text{panel\_size} \cdot\text{GHI}(t)
\end{equation}
}
where \text{panel\_size} is the solar panel size of an MEC server. 
By using $\lambda_{t_1,t_2}$, we can derive the energy arrival between two request arrivals, i.e., $\lambda_{i,i+1}$.
\item \textbf{Simulation of request arrival and data size:}
We use a Poisson process (with arrival rate $\lambda_r$) to generate the request arrival events (similar simulation setting available in \cite{xu2017online,mao2016dynamic}). The data size of the request follows a uniform distribution (similar setting available in \cite{lyu2017optimal,chen2020joint}) in the scale of 10MB to 30MB, i.e., $d_i \sim \text{Uniform}(10,30)$ MB.
\item \textbf{Action space:} We consider 4 possible actions for each request (i.e., three correspond to different levels of frequency and one is rejection action). Formally, for the $i$-th request, $a_i \in \{0,1,2,3\}$, where actions $\{1,2,3\}$ correspond to processing frequency $\{2,3,4\}$GHz and action $0$ induces request rejection.
\item \textbf{Network Structure:} The target network and Q network in NAFA have exactly the same structure: a Deep Neural Network (DNN) model with two hidden layers, each with 200 and 100 neurons, activated by ReLu, and an output layer, which outputs the estimated Q value for all 4 actions. 
\begin{table}[!t]
	\caption{Simulation parameters}
	\label{table simulation}
	\centering
	\begin{tabular}{ccc }
		\hline
		Symbols  & Meanings & Values \\ 
		\hline
		$\nu$ & computation complexity & 2e4 \\
		$\kappa$ & effective switched capacitance & 1e-28 \\
		$m$ & number of CPU cores & 12\\
		$B_{max}$ & battery capacity & 1e6 (Joules)\\
		\text{panel\_size} & solar panel size & 0.5 ($m^2$)\\
		\hline
	\end{tabular}
\end{table}

\item \textbf{Simulation parameters:} All of the simulation parameters have been specified in Table \ref{table simulation}.
\begin{table}[!t]
	\caption{Training hyper-parameters for NAFA}
	\label{table NAFA}
	\centering
	\begin{tabular}{ccc }
		\hline
		Symbols  & Meanings & Values \\ 
		\hline
		$\epsilon_{0}$ & initial exploration factor & 0.5 \\
		$\epsilon_{min}$ & minimum exploration factor & 0.01 \\
		$\xi$ & discount factor for exploration & 3e4\\
		$\gamma$ & learning rate & 5e-4 \\
		$\beta$ & discount factor for rewards  & 0.995\\
		$\zeta$ & target nework update periodicity  & 5000\\
		$|\tilde{X}_i|$ & batch size  & 80\\
		$X_{max}$ & size of replay buffer  & 1e6\\
		\hline
	\end{tabular}
\end{table}
\item \textbf{Hyper-parameters and training details of NAFA:} All hyper-parameters are available in Table \ref{table NAFA}. In our simulation, we use $ep_{max}=30\times 5$ episodes to train the model. The simulation time for each episode of training is 10 straight days. We reset the simulation environment based on different GHI data once the last request within these 10 days has been scheduled.\footnote{So, in our real implementation, training steps $N_{max}$ is not necessarily identical for all the training episodes.} Besides, it is important to note that, since we only have $30\times 10$ days of GHI data as our training dataset, we re-use the same $30\times 10$ days of GHI data for training episodes between $30\times 2$ to $30\times5$.
\item \textbf{Uniformization of states:} In our implementation of NAFA, we have uniformed all the elements' values in a state to the scale of $[0,1]$ before its input to the neural network. We are prone to believe that such a uniformization might potentially improve the training performance.
\end{itemize}
\par
\begin{figure*}[!t]
	\centering
	\includegraphics[ width=6.5in]{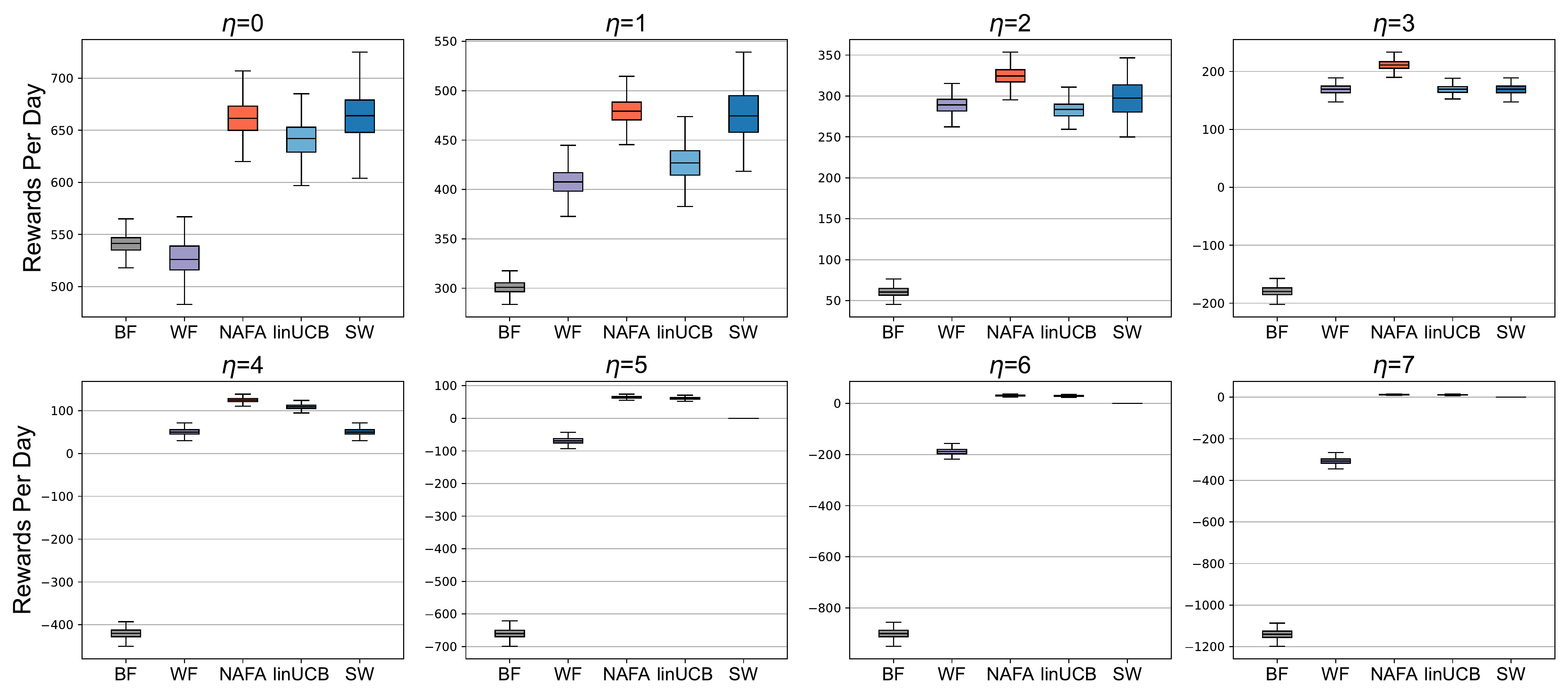}
	\caption{\color{black}Box plot demonstrating rewards per day vs. tradeoff $\eta$ fixing $\lambda_r=30$. Each of the sub-figures demonstrates reward data of 300 straight simulation days, using the same testing dataset.}
	\label{fix arrival rewards}
\end{figure*}
\subsubsection{Baselines}
For an evaluation purpose, we implement three baseline methods, specified as follows:
\begin{itemize}
	\item \textbf{Best Fit (BF):} Best Fit is a rule-based scheduling algorithm derived from \cite{farahnakian2016energy}. In our problem, BF tends to reserve energy for future use by scheduling the minimized processing frequency for the incoming request.
	Explicitly, it selects an action:
	\begin{equation}
		a_i=
	\begin{cases}
		\operatorname{argmin}_{a \in \mathcal{A}_{s_i} } f_{a} & |\mathcal{A}_{s_i}|>1 \\
		0 & \text{otherwise}
	\end{cases} 
	\end{equation}
	
	\item \textbf{Worst Fit (WF):} Worst Fit is another rule-based scheduling algorithm derived from \cite{xian2007energy}.  In our problem, WF is desperate to reduce the processing time of each request. It achieves this goal via scheduling the maximized processing frequency for the incoming request as far as it is possible. Explicitly, it selects an action:
		\begin{equation}
		a_i=
		\begin{cases}
			\operatorname{argmax}_{a \in \mathcal{A}_{s_i} } f_{a} & |\mathcal{A}_{s_i}|>1 \\
			0 & \text{otherwise}
		\end{cases} 
	\end{equation}
	\item \textbf{linUCB:} linUCB is an online learning solution and it is typically applied in a linear contextual bandit model (see \cite{chu2011contextual}). In our setting, linUCB learns the immediate rewards achieved by different contexts (or states in our formulation) and it greedily selects the action that maximizes its estimated immediate rewards. However, it disregards the state transfer probability, or in other words, it ignores the rewards that might be obtained in the future. Besides, same as NAFA, in our implementation of linUCB, we perform the same uniformization process for a state before its training. We do this in a bid to ensure a fair comparison.
	{\color{black}
	\item \textbf{Sliding-Window (SW):} SW is an online learning solution proposed in \cite{liu2020resource}. In our setting, we allow SW to be prophetic, i.e., it is given extra information for the states $s_{i}$, \dots, $s_{i+M}$ while making decision $a_i$. SW always chooses an action that maximizes the cumulative rewards for requests indexed within the window $[i,i+M]$. For each window, it calculates the on-off decision of the first $L \leq M$ requests in the window, utilizing the  request and energy arrival pattern of the entire $M$ requests. In our setting, we set $M=500$ and $L=300$.}
\end{itemize}	
In our experiment, we train linUCB and NAFA for the same amount of episodes. After training, we use 300 straight days of simulation based on the same test dataset to validate the performance of different scheduling strategies.

\subsection{Results}

%

\subsubsection{Tradeoff factor vs. Average Rewards}

With the request arrival rate fixing to $\lambda_r=30$ and different settings of tradeoff factor $\eta$, we shall show how different scheduling strategies work. The results can be viewed in the box plot, shown in Fig. \ref{fix arrival rewards}. From this plot, we can derive the following observations: 
\begin{itemize}
	\item In all the settings of $\eta$, NAFA outperforms all the other baselines in terms of average returned rewards. This result strongly shows the superiority of our proposed solution: not only is NAFA capable of adaptively adjusting its action policy based on the operator's preference (embodied by tradeoff $\eta$), but it also has a stronger learning performance, comparing with another learning algorithm, i.e., linUCB. 
	\item With $\eta$ becoming bigger, NAFA's, linUCB's, and SW's average rewards approach to 0 while other rule-based algorithms (i.e., WF and BF) reduce to a negative number. Our explanation for this phenomenon is that when $\eta$ is set to a sufficiently high value, the penalty brought by processing time has exceeded the real rewards brought by accepting a request (see our definition of rewards in Eq.(\ref{reward})). As such, accepting a request which takes too much time to process, is no longer a beneficial action for this extreme tradeoff setting, so both NAFA and linUCB learn to only accept a small portion of requests (perhaps those with smaller data size) and only acquire a slightly positive reward. {\color{black} The prophetic algorithm SW is given extra rewards information in the future, so it can also adaptively adjust its decision based on changing $\eta$. But the algorithm still cannot receive the largest rewards since it cannot see through the whole training process (i.e., $M$ cannot be large enough to cover the whole training process).  Even though we allow $M$ to be large enough (i.e., give it unconstrained future information), its computation complexity may explode if using deep search to iterate all the action combinations). }
\end{itemize}

\subsubsection{Processing Time vs. Rejection Ratio}
\begin{figure}[!hbtp]
	\centering
	\includegraphics[ width=3.5in]{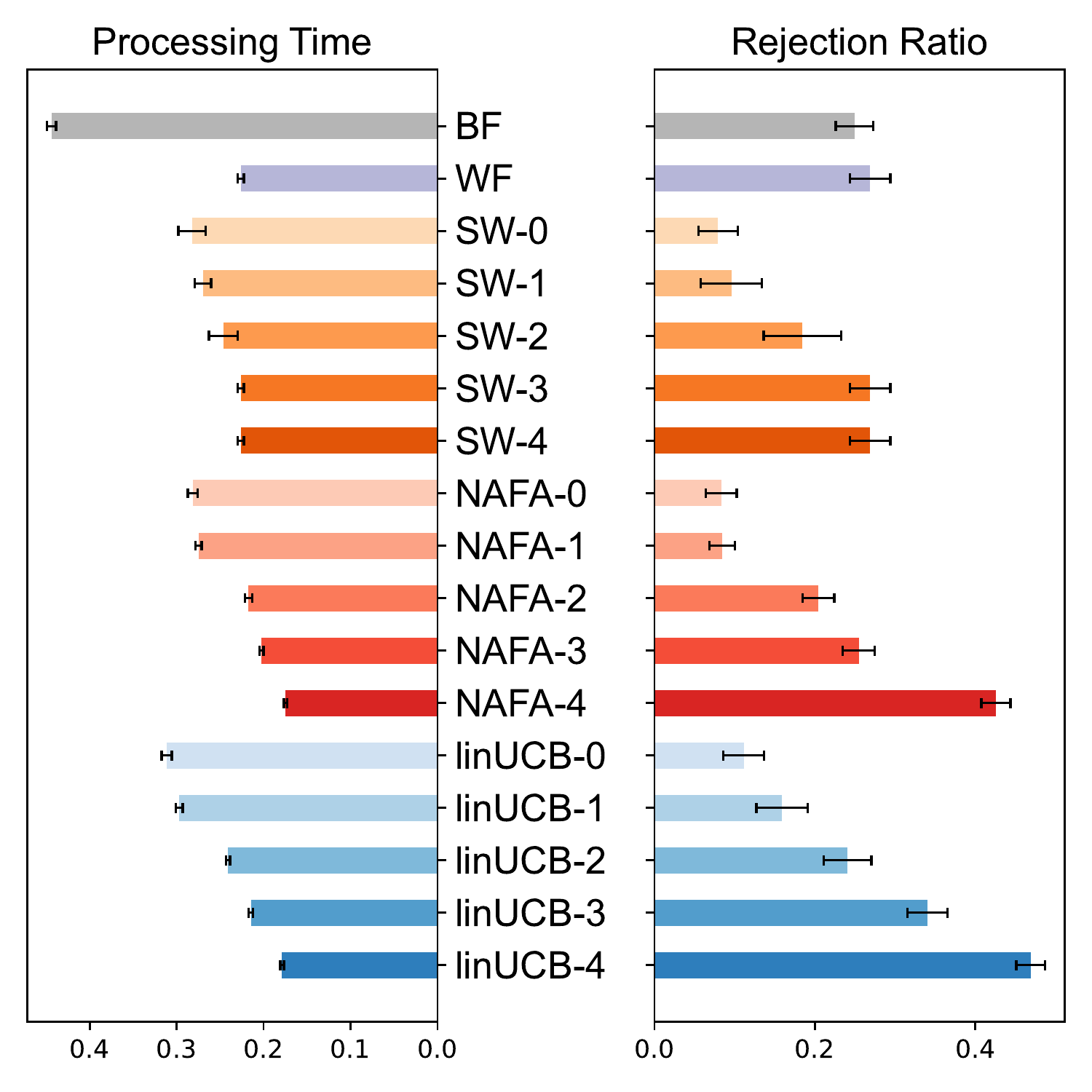}
	\caption{\color{black}Processing time vs. rejection ratio when fixing $\lambda_r=30$. NAFA-\textbf{number}, SW-\textbf{number} and linUCB-\textbf{number} are used to denote different values of $\eta$ setting for different algorithms. }
	\label{latency_fix_lambda}
\end{figure}
Recall that the reward is exactly composed of two parts: real reward of accepting a request, and a penalty of processing time. To draw a clearer picture of these two parts, we show in Fig. \ref{latency_fix_lambda} how different algorithms (and in a different setting of $\eta$) perform in terms of processing time and rejection ratio when fixing $\lambda_r=30$. Intuitively, we derive the following observations:
\begin{itemize}
	\item Comparing BF with WF, BF leads to a higher average processing time, but meanwhile, a lower average rejection ratio is also observed. This phenomenon is wholly comprehensible if considering the working pattern of BF and WF. BF tends to schedule the incoming request to a lower processing frequency in an attempt to conserve energy for future use. This conserved action might lead to a higher average processing time. But at the same time, it is supposed to have a lower rejection ratio when the power supply is limited (e.g. at nights). By contrast, WF might experience more rejection due to power shortage at this time, as a result of its prodigal manner.
	\item With a larger tradeoff $\eta$, NAFA, linUCB and SW all experiences a drop in terms of processing latency, but a rise in the rejection ratio is also observed. It again corroborates that the tradeoff parameter $\eta$ defined in rewards is functioning well, meeting the original design purpose. In this way, operators should be able to adaptively adjust the algorithm's performance based on their own appetites towards the two objectives.
	\item  NAFA-0, NAFA-1, and NAFA-2 significantly outperform BF in terms of both the two objectives, i.e., smaller processing time and lower rejection ratio. Besides, NAFA-2 and NAFA-3 also outperform WF. Of the same tradeoff parameter, linUCB is outperformed by NAFA in terms of both the two objectives in most of the experiment groups. Finally, not an algorithm outperforms NAFA in both the objectives. These observations further justified the superiority of NAFA.
\end{itemize}

\subsubsection{Analysis of Request Treatment and Rejection Motivation}
\begin{figure}[!htbp]
	\centering
	\includegraphics[ width=3.5in]{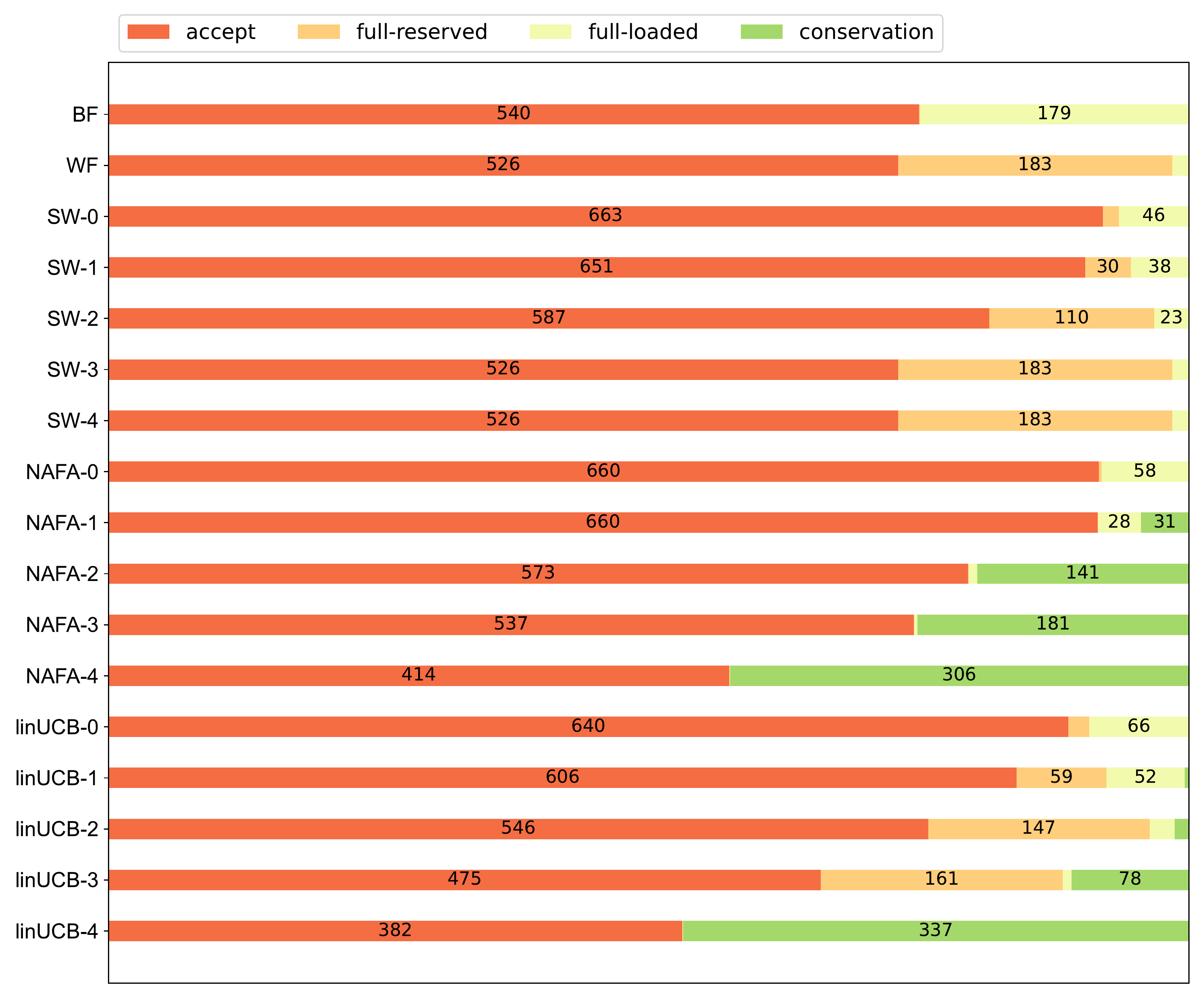}
	\caption{\color{black}Composition of request treatment and rejection motivations for different algorithms when fixing $\lambda_r=30$. We assume the rejections is led by \textbf{full-reserved} when energy in the battery has mostly be reserved, so there is no action can be taken. Analogously, the requests are rejected by the \textbf{full-loaded} when all the free CPU cores are exhausted (in a rare case that the server is simultaneously full-loaded and full-reserved, we count it as full-reserved). Request rejected due to neither of the reasons will be counted as \textbf{conservation} purpose. }
	\label{fix arrival survey}
\end{figure}
\begin{table*}[!tbp]
	\caption{Experimental data under the setting of different tradeoff parameters $\eta$ and a fixed request arrival rate $\lambda_r=30$. {\color{black}Higher acceptance percentage, lower average processing time, and higher rewards are better. Reward is the ultimate goal that we like to evaluate in different settings.} The data of the highest acceptance, the lowest average processing time, and the highest average rewards among the same group of experiment have been highlighted. }
	\label{data fixing lambda}
	\centering
	{\color{black}
		\begin{tabular}{cccccccccc}
			\hline  Tradeoff      & Methods    & \multicolumn{3}{c}{Percentage of Rejection Motivations} &  \textbf{Acceptance}  & \textbf{Average}   & \textbf{Average}  \\
			\cline{3-5} &   &Full-reserved & Full-loaded & Conservation  &  \textbf{Percentage} &\textbf{Processing time} & \textbf{Rewards} & \\
			\hline
			\multirow{4}{*}{$\eta=0$} 
			&BF & 0.06\%& 24.92\%& 0.00\%& 75.02\%& 0.333& 540.957\\
			\cline{2-9}
			&WF& 25.40\%& 1.53\%& 0.00\%& 73.07\%& \textbf{0.165}& 526.897\\ 
			\cline{2-9}
			&NAFA & 0.27\%& 8.05\%& 0.02\%& 91.65\%& 0.258& 660.853\\ 
			\cline{2-9}
			&linUCB & 1.94\%& 9.23\%& 0.00\%& 88.83\%& 0.277& 640.520\\ 
			\cline{2-9}
			&SW& 1.47\%& 6.50\%& 0.00\%& \textbf{92.03}\%& 0.260& \textbf{663.627}\\
			\hline
			\hline
			\multirow{4}{*}{$\eta=2$} 
			&BF& 0.06\%& 24.92\%& 0.00\%& 75.02\%& 0.333& 60.721\\
			\cline{2-9}
			&WF & 25.40\%& 1.53\%& 0.00\%& 73.07\%& \textbf{0.165}& 288.581\\
			\cline{2-9}
			&	NAFA& 0.06\%& 0.78\%& 19.61\%& 79.56\%& 0.173& \textbf{324.494}\\
			\cline{2-9}
			&	linUCB& 20.51\%& 2.33\%& 1.31\%& 75.86\%& 0.183& 283.609\\
			\cline{2-9}
			& SW& 15.27\%& 3.20\%& 0.00\%& \textbf{81.53}\%& 0.201& 297.731\\
			\hline
			\hline
			
			\multirow{4}{*}{$\eta=4$} 
			&BF& 0.06\%& 24.92\%& 0.00\%& \textbf{75.02\%}& 0.333& -419.514\\
			\cline{2-9}
			&WF& 25.40\%& 1.53\%& 0.00\%& 73.07\%& 0.165& 50.266\\
			\cline{2-9}
			&NAFA& 0.06\%& 0.01\%& 42.50\%& 57.44\%& 0.100& \textbf{124.661}\\
			\cline{2-9}
			&linUCB& 0.06\%& 0.02\%& 46.83\%& 53.10\%& \textbf{0.095}& 109.064\\
			\cline{2-9}
			&SW& 25.40\%& 1.53\%& 0.00\%& 73.07\%& 0.165& 50.266\\
			\hline
			\hline
			
			\multirow{4}{*}{$\eta=6$} 
			&BF& 0.06\%& 24.92\%& 0.00\%& \textbf{75.02}\%& 0.333& -899.750\\
			\cline{2-9}
			&WF& 25.40\%& 1.53\%& 0.00\%& 73.07\%& 0.165& -188.050\\
			\cline{2-9}
			&NAFA& 0.06\%& 0.00\%& 74.22\%& 25.73\%& \textbf{0.036}& \textbf{30.253}\\
			\cline{2-9}
			&linUCB& 0.06\%& 0.00\%& 73.90\%& 26.04\%& 0.037& 29.594\\
			\cline{2-9}
			&SW& 0.06\%& 0.00\%& 99.94\%& 0.00\%& 0.000& 0.000\\
			\hline
			\hline
			\multirow{4}{*}{$\eta=8$} 
			&	BF& 0.06\%& 24.92\%& 0.00\%& \textbf{75.02}\%& 0.333& -1379.985\\
			\cline{2-9}
			&	WF& 25.40\%& 1.53\%& 0.00\%& 73.07\%& 0.165& -426.365\\
			\cline{2-9}
			&NAFA& 0.06\%& 0.00\%& 93.58\%& 6.36\%& \textbf{0.008}& \textbf{2.503}\\
			\cline{2-9}
			&linUCB& 0.06\%& 0.00\%& 91.49\%& 8.46\%& 0.010& 2.086\\
			\cline{2-9}
			&SW& 0.06\%& 0.00\%& 99.94\%& 0.00\%& 0.000& 0.000\\
			\hline
			\hline
		\end{tabular}
	}
\end{table*}
To have a closer inspection on the algorithms' performance and to explore the hidden motivation that leads to rejection, by Fig. \ref{fix arrival survey}, we show the composition of requests treatment while again fixing $\eta=30$. Based on the figure, the following observations follow:
\begin{itemize}
	\item Most of the rejections of WF is caused by energy full-reserved while most of the rejections of BF is resulting from resource full-loaded. This phenomenon is in accordance with their respective behavioral pattern. BF, which is thrifty on energy usage, might experience too-long processing time, which in turn leads to unnecessary rejection when all the CPU cores are full (i.e., resource full-loaded). By contrast, a prodigal usage of energy might bring about energy full-reserved, as WF is experiencing.  
	\item  linUCB, NAFA and SW are able to balance the tradeoff between resource full-loaded and energy full-loaded, and thereby, resulting in an increase in the overall request acceptance ratio.
	\item  With $\eta$ becoming larger,  linUCB and NAFA both develop a conservation action pattern: some requests are deliberately rejected when no resource full-loaded or energy full-loaded is experiencing. There are two explanations for this pattern: the first is that a) penalty brought by processing latency exceeds the rewards brought by acceptance, even if the action with the highest frequency is taken. Then there is simply no benefit to accept such a request (typically is a request with a large data size). The second motivation is b) the algorithm learns to reject some specific requests (perhaps those larger ones) if the system is nearly full-loaded and full-reserved. Apparently, the conservation pattern that linUCB develops is mostly based on the first motivation, as it has no regard on the state transfer (or informally, the future), while NAFA should be able to learn both of the two motivations. In addition, corroborated by Fig. \ref{fix arrival rewards}, NAFA could indeed gain us more rewards and therefore substantiate the necessity of full consideration of both the two motivations. SW 
\end{itemize}
To gain us a more accurate observation, we demonstrate the average request treatment data in Table \ref{data fixing lambda}.
\subsubsection{Rewards vs. Request Arrival Rate}
\begin{figure}[!hbtp]
	\centering
	\includegraphics[ width=3.5in]{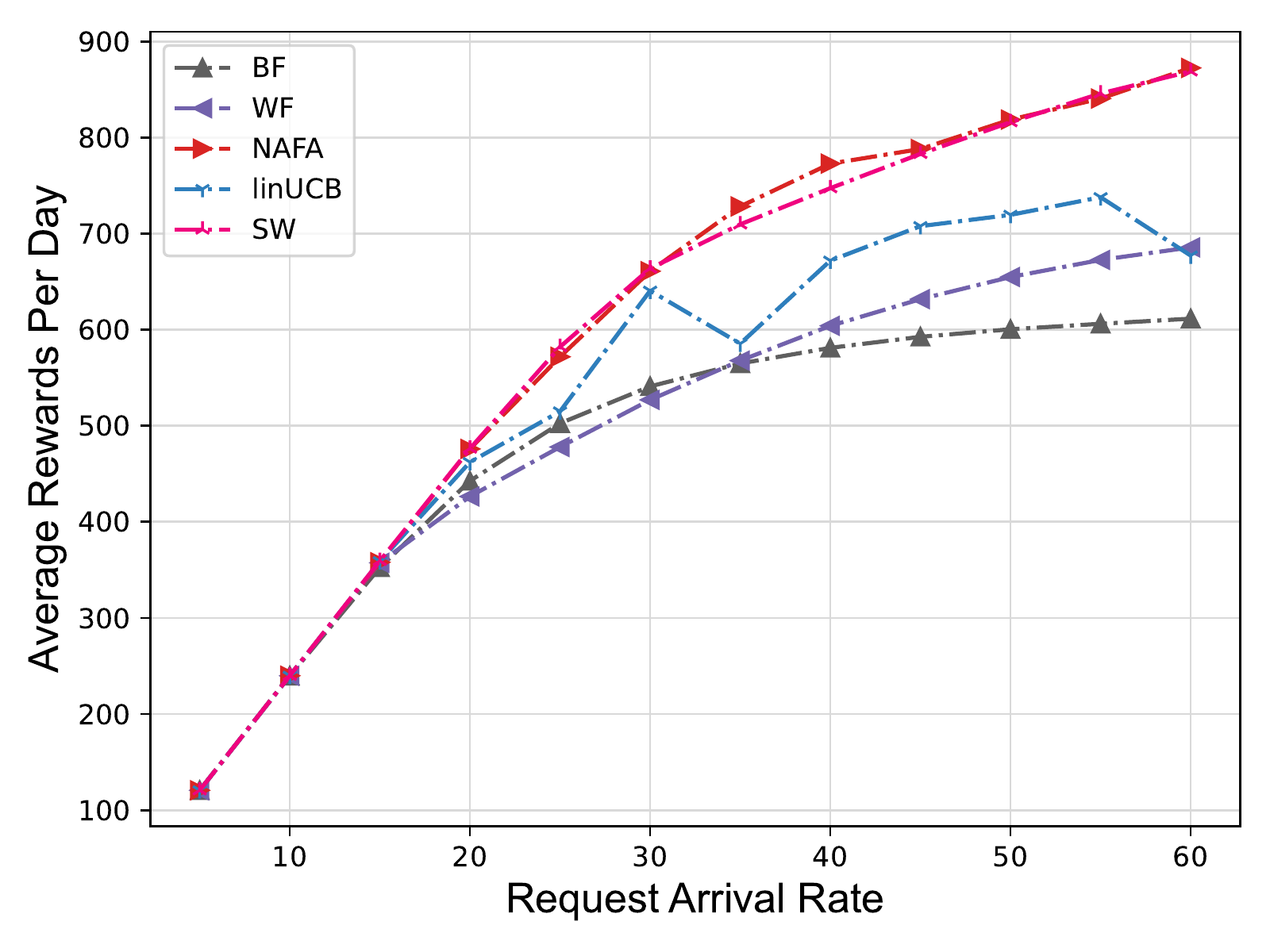}
	\caption{\color{black} Average Rewards vs. Request Arrival Rate when fixing tradeoff parameter $\eta=0$ }
	\label{fix tradeoff=0 rewards}
\end{figure}

\begin{figure}[!hbtp]
	\centering
	\includegraphics[ width=3.5in]{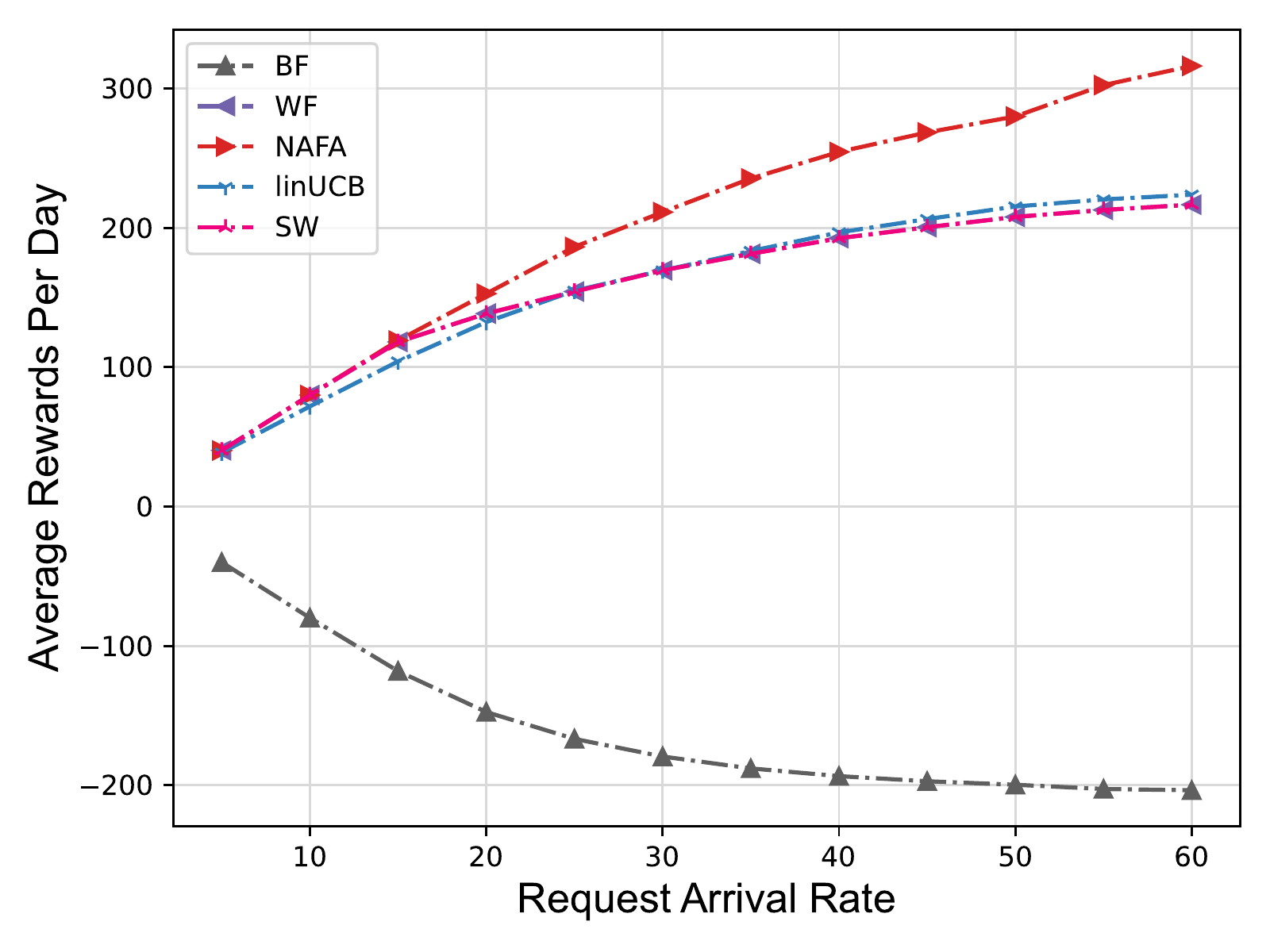}
	\caption{\color{black} Average Rewards vs. Request Arrival Rate when fixing tradeoff parameter $\eta=3$ }
	\label{fix tradeoff=3 rewards}
\end{figure}
In this experiment, we fix the tradeoff parameter to $\eta=0$ and $\eta=3$, and see how the returned rewards evolve with the change of request arrival rate $\lambda_r$. By Fig. \ref{fix tradeoff=0 rewards}, we find that:
\begin{itemize}
	\item Fixing tradeoff $\eta=0$, all the algorithms experience a rise in rewards with the growth of request arrival rate. This phenomenon is quite comprehensible since every single request brings positive rewards (a constant 1 when $\eta=0$) if being proper scheduled.  
	\item However, the growth of WF and BF stagnates as the arrival rate becomes really big. By contrast, NAFA seems to continuously increase in the acquired rewards. By this observation, we see that the traditional rule-based algorithms apparently are incompetent of meeting the scheduling requirement when the system is in high-loaded status.
	\item linUCB experiences a drastic jitter under different request arrival rates. We speculate that this jitter is resulting from a constant reward achieved for every state (or context). When $\eta=0$, the algorithm simply reduces to a random selection since each action has exactly the same expected rewards for almost all the states (except those when the system is full-loaded or full-reserved).
	{\color{black}
	\item SW performs roughly the same with NAFA in the setting of $\eta = 0$. But when $\lambda_r = 35$ and $\lambda_r=40$, a better performance of NAFA is observed.  }
\end{itemize}

By Fig. \ref{fix tradeoff=3 rewards}, we also find that, for $\eta=3$:
\begin{itemize}
\item BF gained a negative average reward. This is resulting from the excessively high processing time penalty if only employing the least processor frequency.
\item NAFA again acquirs the highest average reward in all groups of experiments.

\end{itemize}

\subsubsection{Acceptance Ratio and Processing Time vs. Request Arrival Rate}
\begin{figure}[!hbtp]
	\centering
	\includegraphics[ width=3.5in]{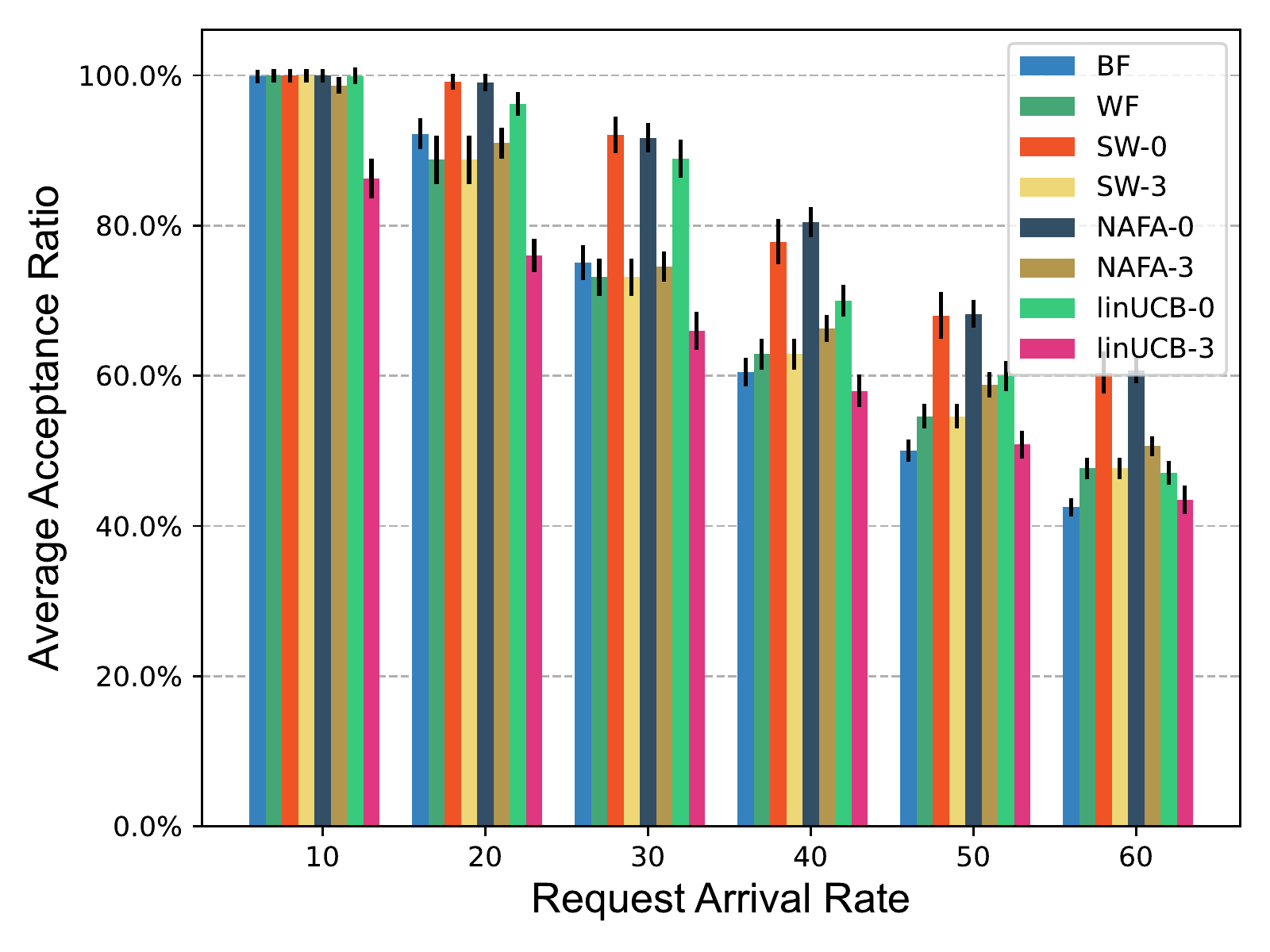}
	\caption{\color{black}Average acceptance ratio vs. request arrival rate. NAFA-\textbf{Number}, SW-\textbf{Number} and linUCB-\textbf{Number} represent the algorithms trained in the setting of $\eta=$\textbf{Number}. }
	\label{acceptance_fix_tradeoff}
\end{figure}
\begin{figure}[!hbtp]
	\centering
	\includegraphics[ width=3.5in]{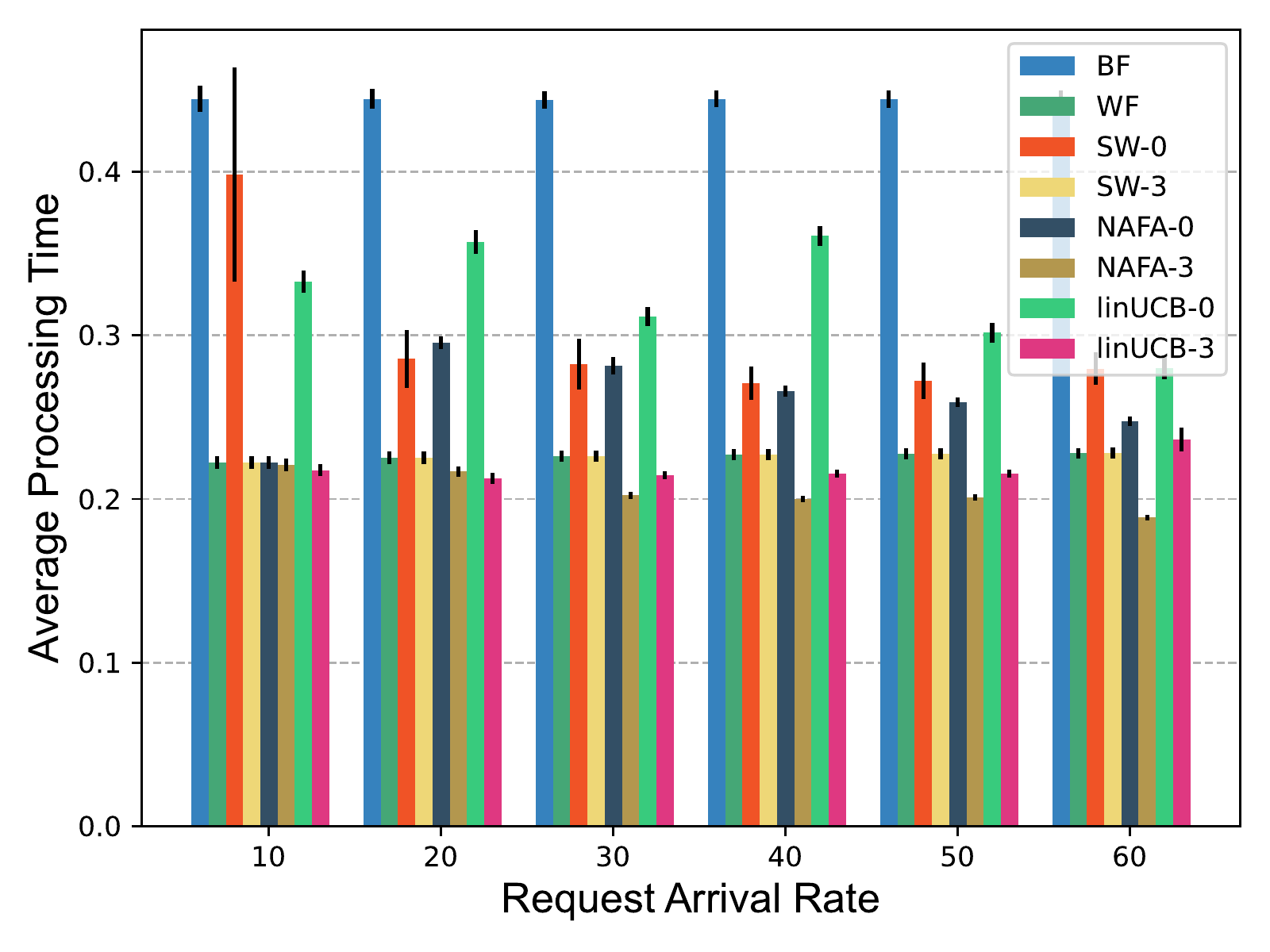}
	\caption{\color{black} Average processing time vs. request arrival rate. NAFA-\textbf{Number}, SW-\textbf{Number} and linUCB-\textbf{Number} represent the algorithms trained in the setting of $\eta=$\textbf{Number}. }
	\label{latency_fix_tradeoff}
\end{figure}
To show the whole picture, we now demonstrate in Fig. \ref{acceptance_fix_tradeoff} and Fig. \ref{latency_fix_tradeoff} how the two objectives, i.e., acceptance ratio and processing time evolve with different request arrival rates. Based on the result, we give the following observations:
\begin{itemize}
	\item Processing time of BF and WF shows little variation with request arrival rate. This phenomenon is intuitive since these rule-based methods are prone to schedule the same frequency to the incoming request (the largest one for WF and the smallest one for BF).
	\item Compared to WF, BF acquires a higher acceptance ratio when the request rate is relatively low, but WF surpasses BF when the request rate becomes greater. Obviously, this phenomenon seems to be telling us that when the request rate is sufficiently high, the system is more vulnerable to resource full-loaded than to energy full-reserved. 
	\item NAFA-0 yields the largest acceptance ratio while NAFA-3 yields the smallest processing time, comparing to other baselines. This corroborates the effectiveness and high adaptiveness of NAFA.
	{\color{black}
	\item SW-0 acquires the second-largest acceptance ratio and linUCB acquires the second-smallest processing time.  This substantiates some sort of adaptiveness of linUCB and SW, However, their performance could not match up with NAFA due to their  "near-sighted" action pattern.}
\end{itemize}
\begin{table*}[!htbp]
	\caption{Experimental data under the setting of different request arrival rates $\lambda_r$, and a fixed tradeoff parameter $\eta=3$. {\color{black}Higher acceptance percentage, lower average processing time, and higher rewards are better. Reward is the ultimate goal we like to evaluate in different settings.} The data of the highest acceptance, the lowest average processing time and the highest average rewards among the same group of experiment have been highlighted.  }
	\label{data with different request rate}
	\centering
	{\color{black}
	\begin{tabular}{cccccccc}
		\hline  Request      & Methods    & \multicolumn{3}{c}{Percentage of Rejection Motivations} &  \textbf{Acceptance}  & \textbf{Average}   & \textbf{Average}  \\
		\cline{3-5}  Arrival Rate &   & Full-reserved & Full-loaded & Conservation  & \textbf{Percentage}&\textbf{Porcessing time} &\textbf{Rewards} \\
		\hline
		\multirow{4}{*}{$\lambda_r=10$}    
	&BF& 0.05\%& 0.01\%& 0.00\%& 99.85\%& 0.444& -79.897\\
	\cline{2-8}
	&WF& 0.05\%& 0.00\%& 0.00\%& \textbf{99.95\%}& 0.222& \textbf{79.931}\\
	\cline{2-8}
	&NAFA& 0.05\%& 0.00\%& 1.28\%& 98.67\%& 0.218& 79.907\\
	\cline{2-8}
	&linUCB& 0.05\%& 0.00\%& 13.70\%& 86.25\%& \textbf{0.188}& 71.819\\
	\cline{2-8}
	&SW& 0.05\%& 0.00\%& 0.00\%& \textbf{99.95}\%& 0.222& \textbf{79.931}\\
	\hline
	\hline
		
		\multirow{4}{*}{$\lambda_r=20$}
&	BF& 0.05\%& 0.78\%& 0.00\%& \textbf{92.12\%}& 0.409& -147.452\\
		\cline{2-8}
&	WF& 11.14\%& 0.01\%& 0.00\%& 88.75\%& 0.200& 138.410\\
		\cline{2-8}
	&NAFA& 0.07\%& 0.01\%& 8.91\%& 90.97\%& 0.197& \textbf{152.735}\\
		\cline{2-8}
	&linUCB& 0.05\%& 0.00\%& 23.90\%& 76.02\%& \textbf{0.162}& 132.316\\
		\cline{2-8}
	&SW& 11.14\%& 0.01\%& 0.00\%& 88.75\%& 0.200& 138.410\\
	\hline
	\hline
		
		\multirow{4}{*}{$\lambda_r=30$}
	&BF& 0.06\%& 2.49\%& 0.00\%& \textbf{75.02\%}& 0.333& -179.397\\
		\cline{2-8}
	&WF& 25.40\%& 0.15\%& 0.00\%& 73.07\%& 0.165& 169.423\\
		\cline{2-8}
	&NAFA& 0.06\%& 0.03\%& 25.12\%& 74.52\%& 0.151& \textbf{211.238}\\
		\cline{2-8}
	&linUCB& 22.40\%& 0.08\%& 10.85\%& 65.93\%& \textbf{0.141}& 169.323\\
		\cline{2-8}
	&SW& 25.40\%& 0.15\%& 0.00\%& 73.07\%& 0.165& 169.423\\
	\hline
	\hline
		
		\multirow{4}{*}{$\lambda_r=40$}
	&BF& 0.05\%& 3.95\%& 0.00\%& 60.49\%& 0.269& -193.469\\
	\cline{2-8}
	&WF& 31.57\%& 0.56\%& 0.00\%& 62.88\%& 0.143& 192.471\\
	\cline{2-8}
	&NAFA& 0.99\%& 0.14\%& 31.34\%& \textbf{66.26\%}& 0.133& \textbf{254.438}\\
	\cline{2-8}
	&linUCB& 29.23\%& 0.37\%& 9.13\%& 57.97\%& \textbf{0.125}& 196.661\\
	\cline{2-8}
	&SW& 31.57\%& 0.56\%& 0.00\%& 62.88\%& 0.143& 192.471\\
	\hline
	\hline
		
		\multirow{4}{*}{$\lambda_r=50$}
	&BF& 0.06\%& 4.99\%& 0.00\%& 50.04\%& 0.222& -199.659\\
	\cline{2-8}
&WF& 34.62\%& 1.08\%& 0.00\%& 54.57\%& 0.124& 207.774\\
	\cline{2-8}
&NAFA& 1.95\%& 0.28\%& 36.46\%& \textbf{58.82\%}& 0.118& \textbf{279.897}\\
	\cline{2-8}
&linUCB& 32.58\%& 0.80\%& 8.57\%& 50.81\%& \textbf{0.110}& 215.257\\
	\cline{2-8}
&SW& 34.53\%& 1.08\%& 0.08\%& 54.58\%& 0.124& 207.791\\
\hline
\hline		
		\multirow{4}{*}{$\lambda_r=60$}
&BF& 0.05\%& 5.75\%& 0.00\%& 42.49\%& 0.189& -203.620\\
\cline{2-8}
&WF& 36.31\%& 1.60\%& 0.00\%& 47.65\%& 0.109& 216.563\\
\cline{2-8}
&NAFA& 1.66\%& 0.23\%& 45.38\%& \textbf{50.62\%}& \textbf{0.096}& \textbf{316.105}\\
\cline{2-8}
&linUCB& 35.18\%& 1.30\%& 8.39\%& 43.47\%& 0.103& 223.750\\
\cline{2-8}
&SW& 36.13\%& 1.60\%& 0.17\%& 47.66\%& 0.109& 216.574\\
		\hline
	\end{tabular}
}
\end{table*}

To enable an accurate observation, we demonstrate in Table \ref{data with different request rate} our experimental data given different request arrival rate and a fixed tradeoff parameter $\eta=3$.

\section{Conclusion and Future Prospect}
In this paper, we have studied an adaptive frequency adjustment problem in the scenario of intermittent energy supplied MEC. Concerning multiple tradeoffs persisted in the formulated problem, we propose a deep reinforcement learning-based solution termed NAFA for problem-solving.  By our real solar data-based experiments, we substantiate the adaptiveness and superiority of NAFA, which drastically outperforms other baselines in terms of acquired reward under different tradeoff settings. \par
For a potential extension of our proposed solution, we are particularly interested in a more advanced technique, known as Federated Learning (FL, \cite{mcmahan2016communication}). FL allows multiple servers to collectively train a Deep Q network, while do not necessarily need to expose their private data to a centralized entity. This novel training paradigm is pretty advantageous over our extensions from the single-server to the multi-servers scenario: we might gain access to request and energy arrival patterns from different servers (perhaps in different venues and owned by different operators) without knowing its training data. {\color{black}Another interesting direction for extension to the multi-servers scenario is to employ meta reinforcement learning (e.g., \cite{qu2021dmro}). By learning properly initialized weights of models for different servers, the reinforcement learning process on different servers might be further accelerated. }



\ifCLASSOPTIONcaptionsoff
  \newpage
\fi



%
\bibliography{NAFA}{}
\bibliographystyle{IEEEtran}
%

%
%
%



\enlargethispage{-5in}
\begin{IEEEbiography}[{\includegraphics[width=1in,height=1.25in,clip,keepaspectratio]{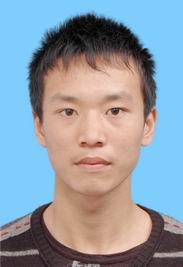}}]{Tiansheng Huang}
 is working towards his M.S. degree with the School of Computer Science and Engineering, South China University of Technology, China. His research interests include fog/edge computing, parallel and distributed computing, and distributed machine learning.
\end{IEEEbiography}
\vfill
\begin{IEEEbiography}[{\includegraphics[width=1in,height=1.25in,clip,keepaspectratio]{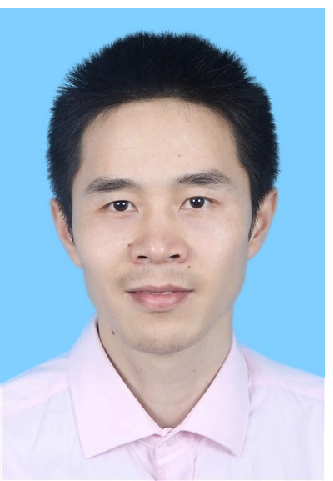}}]{Weiwei Lin}
  received his B.S. and M.S. degrees from Nanchang University in 2001 and 2004, respectively, and the PhD degree in Computer Application from South China University of Technology in 2007. Currently, he is a professor in the School of Computer Science and Engineering, South China University of Technology. His research interests include distributed systems, cloud computing, big data computing and AI application technologies. He has published more than 100 papers in refereed journals and conference proceedings. He has been the reviewers for many international journals, including TPDS, TC, TMC, TCYB, TSC, TCC, etc. He is a senior member of CCF and a member of the IEEE.
\end{IEEEbiography}
\vfill
\begin{IEEEbiography}[{\includegraphics[width=1in,height=1.25in,clip,keepaspectratio]{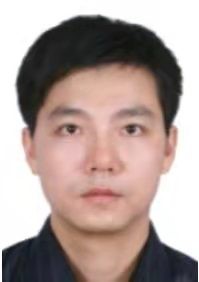}}]{Xiaobin Hong}
 received his Ph.D. degree in mechanical engineering from the South China University of Technology, Guangzhou, in 2007. He is currently a Professor with the School of Mechanical and Automotive Engineering, South China University of Technology. His research interests include instrumentation, signal processing and AI methods.
\end{IEEEbiography}

\begin{IEEEbiography}[{\includegraphics[width=1in,height=1.25in,clip,keepaspectratio]{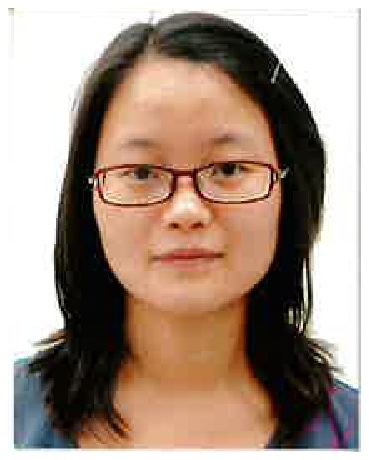}}]{Xiumin Wang}
 is now an Associate Professor in the School of Computer Science and Engineering, South China University of Technology, China. Xiumin Wang received her B.S. from the Department of Computer Science, Anhui Normal University, China, in 2006. She received her Ph.D. degree from both the Department of Computer Science, University of Science and Technology of China, Hefei, China, and the
Department of Computer Science, City University of Hong Kong, under a joint PhD program. Her research interests include mobile computing, algorithm design and optimization.
\end{IEEEbiography}

\begin{IEEEbiography}[{\includegraphics[width=1in,height=1.25in,clip,keepaspectratio]{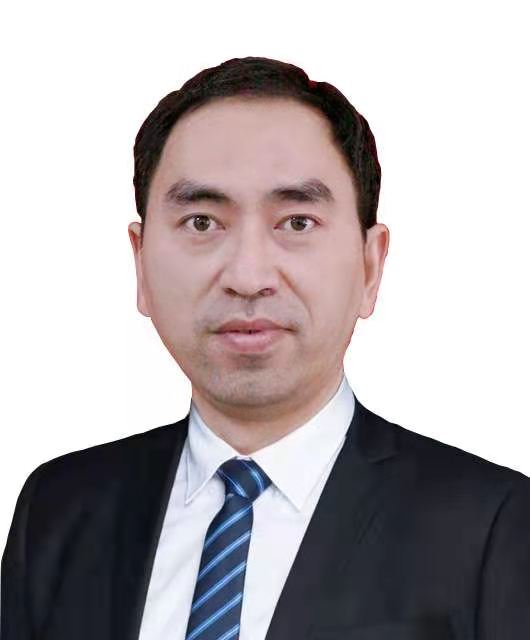}}]{Qingbo Wu}
	received the Ph.D degree in computer science and technology from National University of Defense Technology in 2010. Now he is a professor at National University of Defense Technology. His research interests include operating system and cloud computing.
\end{IEEEbiography}

\begin{IEEEbiography}[{\includegraphics[width=1in,height=1.25in,clip,keepaspectratio]{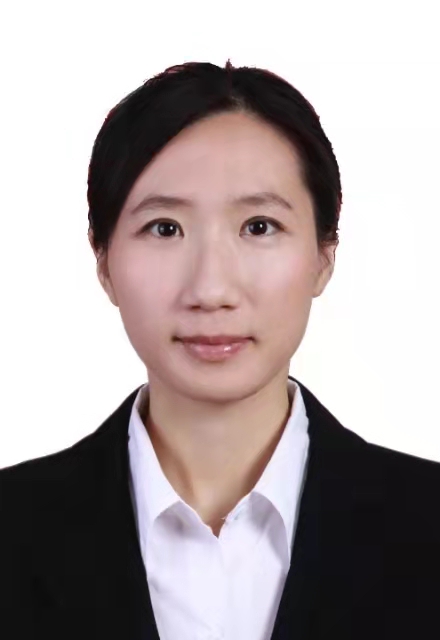}}]{Rui Li}
 received the Ph.D degree in computer software and theory from Beihang University in 2016. Now she is a associate professor at Peng Cheng Laboratory. Her research interests include operational software and cloud computing.
\end{IEEEbiography}

\vfill
\begin{IEEEbiography}[{\includegraphics[width=1in,height=1.25in,clip,keepaspectratio]{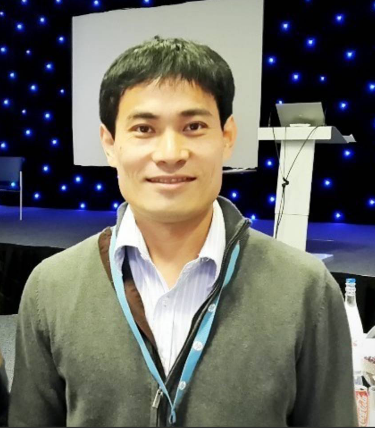}}]{Ching-Hsien Hsu}
 is Chair Professor and Dean of the College of Information and Electrical Engineering, Asia University, Taiwan; His research includes high performance computing, cloud computing, parallel and distributed systems, big data analytics, ubiquitous/pervasive computing and intelligence. He has published 200 papers in top journals such as IEEE TPDS, IEEE TSC, ACM TOMM, IEEE TCC , IEEE TETC, IEEE System, IEEE Network, top conference proceedings, and book chapters in these areas. Dr. Hsu is the editor-in-chief of International Journal of Grid and High Performance Computing, and International Journal of Big Data Intelligence; and serving as editorial board for a number of prestigious journals, including IEEE Transactions on Service Computing, IEEE Transactions on Cloud Computing, International Journal of Communication Systems, International Journal of Computational Science, AutoSoft Journal. He has been acting as an author/co-author or an editor/co-editor of 10 books from Elsevier, Springer, IGI Global, World Scientific and McGraw-Hill. Dr. Hsu was awarded six times talent awards from Ministry of Science and Technology, Ministry of Education, and nine times distinguished award for excellence in research from Chung Hua University, Taiwan. Since 2008, he has been serving as executive committee of IEEE Technical Committee of Scalable Computing; IEEE Special Technical Co mmittee Cloud Computing; Taiwan Association of Cloud Computing. Dr. Hsu is a Fellow of the IET (IEE); Vice Chair of IEEE Technical Committee on Cloud Computing (TCCLD), IEEE Technical Committee on Scalable Computing (TCSC), a Senior member of IEEE.
\end{IEEEbiography}
\vfill
\begin{IEEEbiography}[{\includegraphics[width=1in,height=1.25in,clip,keepaspectratio]{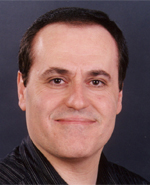}}]{Albert Y. Zomaya} is currently the chair professor of High Performance Computing \& Networking in the School of Information Technologies, The University of Sydney. He is also the director in the Centre for Distributed and High Performance Computing, which was established in late 2009. He published more than 500 scientific papers and articles and is a author, co-author or editor of more than 20 books. He served as the editor in chief of the IEEE Transactions on Computers (2011-2014). He serves as an associate editor for 22 leading journals, such as, the ACM Computing Surveys, IEEE Transactions on Computational Social Systems, IEEE Transactions on Cloud Computing, and Journal of Parallel and Distributed Computing. He delivered more than 150 keynote addresses, invited seminars, and media briefings and has been actively involved, in a variety of capacities, in the organization of more than 600 national and international conferences. He received the IEEE Technical Committee on Parallel Processing Outstanding Service Award (2011), the IEEE Technical Committee on Scalable Computing Medal for Excellence in Scalable Computing (2011), and the IEEE Computer Society Technical Achievement Award (2014). He is a chartered engineer, a fellow of AAAS, IEEE, and IET (United Kingdom). His research interests are in the areas of parallel and distributed computing and complex systems. He is a fellow of the IEEE.
\end{IEEEbiography}

{\color{black}
	\appendices
	\section{Justification of a Constant State Transferred Probability}
	\label{appendix A}
	In our methodology, we apply reinforcement learning technique to learn the state transferred probability, which is assumed to be a constant given current state and action, i.e.,   
	\begin{equation}
		\label{state transfer probability2}
		p(s_{i+1} |s_i,a_i)= \text{a constant}
	\end{equation}
	To make the above assumption true, the state formulation has to be careful constructed, so as to maintain the numerical stability. In what follows, we shall give several observations to show the motivation behind our state formulation, and tell under what assumptions can we make sure the above property holds true.  
	
	Explicitly, Equation (\ref{state transfer probability2}) tells we need to make sure that the corresponding elements (e.g., $B_{i+1}$, $S_{i+1}$ and $\psi_{i+1,1} \dots, \psi_{i+1,n}$) in state $s_{i+1}$ follows the same joint \textit{stationary distribution} by given the same $s_i$. A \textit{stationary distribution} means it is no longer relevent with request order $i$ if given $s_i$. To show this desirable property, we first need to show the following observation.
\begin{observation}[Stationary Energy Arrival \textbf{Given State}]
	\label{stationary energy arrival}
	By specifying $T_i$, the captured energy between two sequential requests (i.e., $\lambda_{i,i+1}$) can be roughly regarded as samples from a stationoary distribution, i.e., value of r.v. $\lambda_{i,i+1}$ is not relevant with $i$ (order of request) given $T_i$.
\end{observation} 
\begin{remark}
	This observation is common, as most intermittent energy supply sources, such as solar power, wind power, have a clear diurnal pattern, but if we fix the time to a specific timestamp in a day, the energy harvest can roughly be regarded as samples from a fixed distribution. Informally, we assume a fixed energy harvest rate for a fixed timestamp, e.g., 1000 Joules/s harvest rate in 3:00 pm. Given request arrival time $T_i$, e.g., 3:00 pm in our example, harvest energy between two events i.e., $\lambda_{i,i+1}$ is stationary. {\color{black} In this way, we respond to the non-stationary issue in challenge 2), \textbf{unknown and non-stationary supplied pattern of energy}, that we previously proposed below P2.} Similarly, one can easily extend the state formulation by covering more system status (e.g., the location of the MEC server, if we aim to train a general model for multiple MEC servers scenario) to yield a more accurate estimation.  
\end{remark}
\begin{observation}[Stationary Request Arrival \textbf{Given State}]
	\label{stationary request arrival}
	By specifying $T_i$, the time interval between request arrival should also be considered as samples from a stationary distribution (though still unknown). This means that the time interval between two arrivals (denoted by $t_{i,i+1}$) is no longer relevant with $i$ if given $T_i$.
\end{observation}
\begin{remark}
	{\color{black}
		By this state formulation, we address the non-stationary issue in challenge 3), \textbf{unknown and non-stationary arrival pattern}, that we previously proposed below P2.}
\end{remark}
\begin{observation}[Stationary Energy Consumption \textbf{Given State}]
	\label{stationary charged energy}
	By specifying $a_i$, $\psi_{i,1} \dots, \psi_{i,n}$ and a stationary time interval between two arrivals, the consumed energy between two requests (i.e., $\rho_{i, i+1}$) could roughly be regarded as samples from a stationary distribution.  
\end{observation}
\begin{remark}
	Our explanation for observation \ref{stationary charged energy} is that the consumed energy $\rho_{i, i+1}$ is actually determined by current CPU frequency as well as the time interval between requests (i.e., $t_{i,i+1}$). Explicitly, we can roughly estimate that $\rho_{i, i+1}=\sum_{n^{\prime}=1}^n  (\psi_{i, n^{\prime}}+ \mathbb{I}\{ a_i= n^{\prime}\}) t_{i,i+1}$. But this calculation is not 100\% accurate since the CPU could have sleeped before arrival of next request, as a reuslt of the process finishing of a request. As a  refinement, we further assume $\rho_{i, i+1}=\sum_{n^{\prime}=1}^n  (\psi_{i, n^{\prime}}+ \mathbb{I}\{ a_i= n^{\prime}\})\cdot t_{i,i+1}- y(a_i, \psi_{i, 1},\dots,\psi_{i, n}, t_{i,i+1})$ where  $y(a_i, \psi_{i, 1},\dots,\psi_{i, n}, t_{i,i+1})$ is a stochastic noise led by the halfway sleeping of CPU cores. Without loss of generality, $y(a_i, \psi_{i, 1},\dots,\psi_{i, n}, t_{i,i+1})$ is assumed to be samples from a stationary distribution given $a_i, \psi_{i, 1},\dots,\psi_{i, n}$ and $t_{i,i+1}$. By this assumption, we state that $\rho_{i, i+1}$ is stationary if given  $a_i, \psi_{i, 1},\dots,\psi_{i, n}$ and a stationary $t_{i,i+1}$.
\end{remark}
\begin{observation}[Stationary Battery and Reserved Status \textbf{Given State}]
	\label{stationary B and S}
	By specifying $s_i$ and $a_i$, $B_{i+1}$ and $S_{i+1}$  can roughly be regarded as samples from two stationary distributions.
\end{observation}
\begin{remark}
	See Eqs. (\ref{battery status}) and (\ref{energy reservation status}), we find that $B_{i+1}$ and $S_{i+1}$ are relevant with $\lambda_{i,i+1}$, $\rho_{i,i+1}$, $e_{a_i}$ and $s_i$. As per Observation \ref{stationary energy arrival} and \ref{stationary charged energy}, $\lambda_{i,i+1}$ and $\rho_{i,i+1}$ are all stationary given $s_i$ and $a_i$. In addition, we note that $e_{a_i}$ is stationary given the same condition since  $d_i$ is assumed to be samples from a stationary distribution and there is not other stochastic factor in  Eq. (\ref{energy consumption}). 
\end{remark}
\begin{observation}[Stationary Core Status \textbf{Given State}]
	\label{stationary core frequency}
	By specifying $s_i$ and $a_i$,  $\psi_{i+1, 1},\dots,\psi_{i+1, n}$ can roughly be regarded as samples from a stationary distribution.
\end{observation}
\begin{remark}
	Here we simply regard that $\psi_{i+1, n}$ =  $\psi_{i, n}+ \mathbb{I}\{ a_i= n^{\prime}\}- z(\psi_{i, n}, t_{i,i+1})$ where  $z(\psi_{i, n}, t_{i,i+1})$ denotes the reduction on number of active core in $f_n$ GHz during $ t_{i,i+1}$ interval. Instinctively, we feel that the reduction, i.e., $z(\psi_{i, n}, t_{i,i+1})$ should at least have some sort of bearing with the running core status, i.e., $\psi_{i, n}$ and the time interval between requests i.e., $t_{i,i+1}$. Informally, we might simply imagine that for each time unit, the amount of $\alpha_n \cdot \psi_{i, n} $ expected reduction would be possibly incurred, where $\alpha_n$ should be only relevant to the evaluated core frequency (i.e., $f_n$).  Then, following this strand of reasoning, $z(\psi_{i, n}, t_{i,i+1})$ should be a stochastic variable and it can be roughly regarded as samples from a stationary distribution if given $\psi_{i, n}$ and $t_{i,i+1}$.
\end{remark}
\begin{observation}[Stationary Time and Data Size \textbf{Given State}]
	\label{chore element}
	By specifying $s_i$ and $t_{i,i+1}$ , the timestamp in which the next request comes, i.e., $T_{i+1}= T_{i}+t_{i,i+1}$ is naturally stationary. Besides, the data size of a request, i.e., $d_{i+1}$ is naturally assumed to be samples from a stationary distribution (i.e., not relevant with the arrival order $i$).
\end{observation}
Combining Observation   \ref{stationary B and S}, \ref{stationary core frequency} and \ref{chore element}, we roughly infer that there should be a unique constant specifying the transferred probability from a state to another. However, we are aware that this conclusion is not formal in a very rigorous sense, given the complexity that exists in the analysis as well as the \textbf{not necessarily 100\% correct} assumptions we make in the observation. \par
	But in practice, the assumption does not necessarily need to be 100\% rigorous, we only need to construct a state formulation that let $p(s_{i+1} |s_i,a_i)  \approx \text{a constant}$, but with less noise as far as possible. Our real intention of discussing the stationary property is actually to render the readers an instruction about state formulation, and show that how to extend the state if in a different application case.
}

%
{\color{black}
	\section{Example illustrating SMDP Formulation}
	\label{appendix B}
	\begin{figure*}[!hbtp]
		\centering
		\includegraphics[width=6.5in]{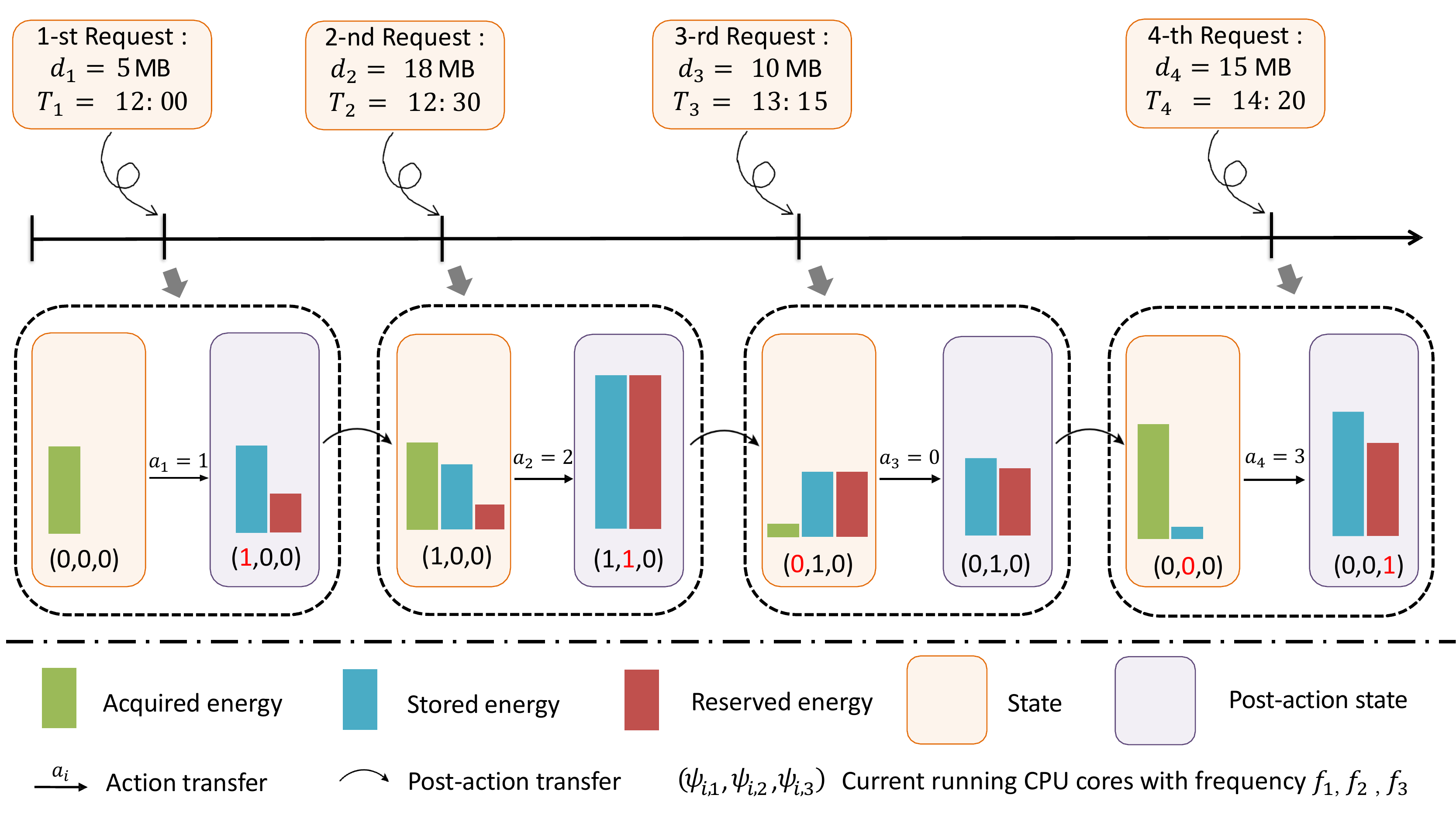}
		\caption{Illustration of the event-driven SMDP model }
		\label{event demonstration}
	\end{figure*}
	In this sub-section,  we shall show the readers a simplified working procedure of our proposed SMDP model. As shown in Fig. \ref{event demonstration}, we consider three possible frequency adjustment options for each incoming request, i.e., $a_i \in \{0, 1, 2, 3\}$ where action $a_i=0$ means rejection and actions $a_i=1, 2, 3$ respectively correspond to a frequency setting of $f_1, f_2, f_3$. Our example basically illustrates the following workflows:
	\begin{itemize}
		\item   Before the 1-st request's arrival, a specific amount of energy (i.e., \textbf{acquired energy}, represented by the green bar) has been collected by the energy harvest modules. Then, at the instance of the 1-st request's arrival, the scheduler decides to take action $a_1=1$ based on the current state \footnote{Note that in our formal state formulation (see Eq. (\ref{state formulation})), we do not distinguish between \textbf{acquired energy} and \textbf{stored energy}, but only interest in the \textbf{battery status}, which is the sum of these two terms. We make this distinction in this example mainly to support our illustration of the key idea. }. Once the action is being taken, a) a sleeping core would waken and started to run in the frequency of $f_1$, b) a specific amount of energy would be reserved for processing of this request,  c) and the acquired energy would be officially deemed as \textbf{stored energy}. Correspondingly,  the state instantly transfers to a virtual \textbf{post-action state}, i.e., a) the first bit in core status (i.e. running core in the frequency of $f_1$) would be correspondingly flipped to 1, b) the reserved energy would be updated, c) and the newly acquired energy would be absorbed into the stored energy.
		\item  Then, the state (i.e., the system status) continues to evolve over time: the running cores continuously consume the reserved energy and the stored energy (same amount would be consumed for reserved energy and stored energy). The evolvement of state pauses when the second request arrives. After observing the current system status (i.e., current state), the scheduler makes an action $a_2 =2$ at this time and the state similarly transfers to the post-action state.
		\item  Again, after the action being taken, the state evolves over time, and within this time interval, a core (in frequency $f_1$) has finished its task. Predictably, when the 3-rd request arrives, a state with core status $(0,1,0)$ would be observed, and since the available energy (sum of acquired energy and stored energy) is not sufficient for processing this request, the scheduler has no choice but to reject the request,  which makes $a_3=0$. This time, the post-action state does not experience substantial change compared with the observed state (except that acquired energy has been absorbed).
		\item The same evolvement continues and the same action and state transformation process would be repeated for the later requests.
	\end{itemize}
	Please note that in our former formulation, we do not involve the consideration of post-action state, since we only care about the state transformation between two formal states $s_{i}$ and $s_{i+1}$.  We present the concept of post-action state here mainly in a bid to render the readers a whole picture about how the system states might evolve over time.

\section{Detailed Explanation for Discounted State-Action Value}
\label{Appendix C}
Recall that our ultimate goal is to maximize the expected cumulative reward, which means that we need to find a deterministic \textit{optimal policy} $ \tilde{\pi}^*$ such that:
\begin{equation}
	\label{ultimate target}
	\tilde{\pi}^*= \mathop{\arg\max}_{a \in \mathcal{A}_s} \tilde{Q}_{\tilde{\pi}^*}(s,a)
\end{equation}
where 
\begin{equation}
	\begin{split}
		&\tilde{Q}_{\tilde{\pi}}(s, a) \\ 
		=&  \mathbb{E}_{s_2,s_3,\dots}  \left[   r(s_1,a_1)+   \sum_{i=2}^{\infty}   r(s_i, \tilde{\pi}(s_i))  | s_1=s ,a_1=a   \right]
	\end{split}
\end{equation}
represents the expected cumulative rewards, starting from an initial state $s$ and an initial action $a$. $\tilde{\pi}^*$ is a deterministic policy that promises us the best action in cumulating expected rewards. However, it is intuitive to find that $\tilde{Q}_{\tilde{\pi}}(s,a)$ is not a convergence value no matter how the policy $\pi$ is defined (see the summation function), which makes it meaningless to derive $\pi^*$ in this form. \par
To address this issue, we alternatively define a \textit{discounted expected cumulative rewards}, in the following form:
\begin{equation}
	\begin{split}
		\label{Q function}
		&Q_{\pi}(s,a) \\ 
		=&  \mathbb{E}_{s_2,s_3,\dots}  \left[   r(s_1,a_1)+  \sum_{i=2}^\infty   \beta^{i-1} r(s_i, \pi(s_i))  | s_1=s ,a_1=a   \right]\\
	\end{split}
\end{equation}
where $0<\beta<1$ is the discount factor and $s$ is an initial state.
\par
And we instead  need to find a \textit{near-optimal policy} $\pi^*$ such that:
\begin{equation}
	\begin{split}
		\label{transferred target}
		\pi^*(s)=  \mathop{\arg\max}_{a \in \mathcal{A}_s}  Q_{\pi^*}(s,a)
	\end{split}
\end{equation}
where $s$ is an initial state.  $Q_{\pi}(s,a)$ is usually referred to as \textit{state-action value function} and each $Q_{\pi}(s,a)$ regarding different policy $\pi$  has a finite convergence value, which make it concrete to find $Q_{\pi^*}(s,a)$. And as long as we know about $Q_{\pi^*}(s,a)$, we are allowed to derive the near-optimal policy $\pi^*$ by Eq. (\ref{transferred target}). Moreover,   $Q_{\pi}(s,a)$ still conserves certain information, even after our discount of future reward: we discount much to the reward that might be obtained in the distant future but not that much to the near one, so it actually partially reveals the exact value of a station-action pair (i.e., the cumulative rewards that might gain in the future). Actually, we can state that  $ \pi^* \approx \tilde{\pi}^*  $. Later in our analysis, we would alternatively search for $ \pi^*$ as our target. \par
To derive $Q_{\pi^*}(s,a)$, we need to notice that: 
\begin{equation}
	\begin{split}
		\label{Q value 2}
		Q_{\pi^*}(s, \pi^*(s)  )= \max_{a \in \mathcal{A}_s}  Q_{\pi^*}(s,a)
	\end{split}
\end{equation}
The above result follows our definition of $\pi^*$ in Eq. (\ref{transferred target}). \par
Besides, as per Eq.  (\ref{Q function}), $Q_{\pi}(s,a)$ can indeed rewrite to the following form:
\begin{equation}
	\begin{split}
		\label{Q value 3}
		&Q_{\pi}(s,a) \\ 
		=&  \mathbb{E}_{s_2}  \left[   r(s_1,a_1)+  \beta Q_{\pi}(s_2,\pi (s_2))  | s_1=s ,a_1=a   \right]
	\end{split}
\end{equation}
Plugging $\pi^*$ into Eq. (\ref{Q value 3}), it yields:
\begin{equation}
	\begin{split}
		\label{optimal Q function}
		&Q_{\pi^*}(s,a) \\ 
		=&  \mathbb{E}_{s_2}  \left[   r(s_1,a_1)+  \beta Q_{\pi^*}(s_2,\pi^* (s_2))  | s_1=s ,a_1=a   \right]
	\end{split}
\end{equation}
And plugging Eq. (\ref{Q value 2}) into Eq. (\ref{optimal Q function}) , we have:
\begin{equation}
	\begin{split}
		\label{iterate update function}
		&Q_{\pi^*}(s,a) \\ 
		=&  \mathbb{E}_{s_2}  \left[   r(s_1,a_1)+  \beta  \max_{a_2 \in \mathcal{A}_{s_2}}  Q_{\pi^*}(s_2,a_2)    | s_1=s ,a_1=a   \right]
	\end{split}
\end{equation}
As per the \textbf{Markov} and \textbf{temporally homogeneous} property of an SMDP,  we have:
\begin{equation}
	\begin{split}
		\label{final iterate update function}
		&Q_{\pi^*}(s,a) =  \mathbb{E}_{s^{\prime}}  \left[   r(s,a)+  \beta  \max_{a^{\prime} \in \mathcal{A}_{s^{\prime}}}  Q_{\pi^*}(s^{\prime},a^{\prime})      \right]
	\end{split}
\end{equation}
where $s^{\prime}$ is a random variable (r.v.), which represents the next state that current state $s$ will transfer to. This reach the Bellman optimality that we use in order to learn $Q_{\pi^*}(s,a)$.
}
\end{document}